\documentclass[aps,twocolumn,showpacs,preprintnumbers,floats]{revtex4}\def\@cite#1#2{\textsuperscript{[{#1\if@tempswa , #2\fi}]}}

\usepackage{graphicx}
\usepackage{amsmath}
\usepackage{amsfonts}
\usepackage{amssymb}
\usepackage{color}
\usepackage{subfigure}
\usepackage{epsfig}
\usepackage{morefloats}
\usepackage{multirow}
\usepackage{graphicx,booktabs}
\usepackage{mathrsfs}
\usepackage{txfonts}
\usepackage{indentfirst}
\usepackage{graphicx,booktabs}

\usepackage{longtable,lscape}
\usepackage[colorlinks, citecolor=blue,anchorcolor=red,menucolor=red, linkcolor=red,filecolor=red,urlcolor=blue,frenchlinks=red]{hyperref}

\newcommand{\vlab}{\mbox{\boldmath$\lambda$\unboldmath}}

\newcommand{\vxi}{\mbox{\boldmath$\xi$\unboldmath}}

\begin{document}

\title{Fully-strange tetraquark $ss\bar{s}\bar{s}$ spectrum and possible experimental evidence }

\author{Feng-Xiao Liu$^{1,4}$, Ming-Sheng Liu$^{1,4}$,
Xian-Hui Zhong$^{1,4}$~\footnote {E-mail: zhongxh@hunnu.edu.cn}, Qiang Zhao$^{2,3,4}$~\footnote {E-mail: zhaoq@ihep.ac.cn}}

\affiliation{ 1) Department of Physics, Hunan Normal University, and Key Laboratory of Low-Dimensional Quantum Structures and Quantum Control of Ministry of Education, Changsha 410081, China }

\affiliation{ 2) Institute of High Energy Physics, Chinese Academy of Sciences, Beijing 100049, China}

\affiliation{ 3) University of Chinese Academy of Sciences, Beijing 100049, China}

\affiliation{ 4)  Synergetic Innovation Center for Quantum Effects and Applications (SICQEA),
Hunan Normal University, Changsha 410081, China}
%

%\date{\today}

\begin{abstract}

In this work we construct 36 tetraquark configurations for the $1S$-, $1P$-, and $2S$-wave states, and make a prediction of the mass spectrum for the tetraquark $ss\bar{s}\bar{s}$ system in the framework of a nonrelativistic potential quark model without the diquark-antidiquark approximation. The model parameters are well determined by our previous study of the strangeonium spectrum. We find that the resonances $f_0(2200)$ and $f_2(2340)$ may favor the assignments of ground states
$T_{(ss\bar{s}\bar{s})0^{++}}(2218)$ and $T_{(ss\bar{s}\bar{s})2^{++}}(2378)$, respectively, and the newly observed $X(2500)$ at BESIII may be a candidate of the lowest mass $1P$-wave $0^{-+}$ state $T_{(ss\bar{s}\bar{s})0^{-+}}(2481)$. Signals for the other $0^{++}$ ground state
$T_{(ss\bar{s}\bar{s})0^{++}}(2440)$ may also have been observed in the $\phi\phi$ invariant mass spectrum in $J/\psi\to\gamma\phi\phi$ at BESIII.
The masses of the $J^{PC}=1^{--}$ $T_{ss\bar{s}\bar{s}}$ states are predicted to be in the range of $\sim 2.44-2.99$ GeV,
which indicates that the $\phi(2170)$ resonance may not be a good candidate of the $T_{ss\bar{s}\bar{s}}$ state.
This study may provide a useful guidance for searching for the $T_{ss\bar{s}\bar{s}}$ states in experiments.

\end{abstract}


\maketitle

\section{Introduction}

From the Review of Particle Physics (RPP) of Particle Data Group~\cite{Tanabashi:2018oca},
above the mass range of $2.0$ GeV one can see that there are several unflavored $q\bar{q}$
isoscaler states, such as $f_0(2200)$, $f_2(2150)$, $f_2(2300)$, $f_2(2340)$ etc.,
dominantly decaying into $\phi\phi$, $\eta\eta$, and/or $K\bar{K}$ final states.
The decay modes indicate that these states might be good candidates for conventional
$s\bar{s}$ meson resonances. Recently, we carried out a systematical
study of the mass spectrum and strong decay properties of the $s\bar{s}$ system in Ref.~\cite{Li:2020xzs}.
It shows that these states cannot be easily accommodated by the conventional
$s\bar{s}$ meson spectrum. While they may be candidates for tetraquark $ss\bar{s}\bar{s}$ ($T_{ss\bar{s}\bar{s}}$) states, it is easy to understand that they can fall apart into $\phi\phi$ and $\eta\eta$ final states through quark rearrangements, or easily decay into $K\bar{K}$ final states through a pair of $s\bar{s}$ annihilations and then a pair of light quark creations.
The mass analysis with the relativistic quark model in Ref.~\cite{Ebert:2008id}
supports the $f_0(2200)$, and $f_2(2340)$ to be assigned as the $T_{ss\bar{s}\bar{s}}$ ground states with $0^{++}$ and $2^{++}$,
respectively. However, a relativized quark model calculation~\cite{Lu:2019ira} only favors $f_2(2300)$ to be a $T_{ss\bar{s}\bar{s}}$ state.

Some other candidates of the $T_{ss\bar{s}\bar{s}}$ states from experiment are also suggested
in the literature. For example the vector meson resonance $\phi(2170)$ listed in RPP~\cite{Tanabashi:2018oca}
is suggested to be a $1^{--}$ $T_{ss\bar{s}\bar{s}}$ state based on the mass analysis of QCD sum rules~\cite{tetra1,tetra2,tetra3,tetra5,tetra7}, and flux-tube model~\cite{tetra6}.
The newly observed $X(2239)$ resonance in the $e^+e^-\to K^+K^-$ process at BESIII~\cite{Ablikim:2018iyx} is suggested to be a candidate of the lowest $1^{--}$ $T_{ss\bar{s}\bar{s}}$ state in a relativized quark model~\cite{Lu:2019ira}. Moreover, the newly observed resonances $X(2500)$ observed in $J/\psi\to \gamma \phi\phi$~\cite{Ablikim:2016hlu} and $X(2060)$ observed in $J/\psi\to \phi \eta \eta'$~\cite{Ablikim:2018xuz} at BESIII are suggested to be $0^{-+}$ and $1^{+-}$ $T_{ss\bar{s}\bar{s}}$ states, respectively, according to the QCD sum rule studies~\cite{Dong:2020okt,Cui:2019roq}. The assignment of $X(2500)$ is consistent with that in Ref.~\cite{Lu:2019ira}.

With the recent experimental progresses more quantitative studies on the $T_{ss\bar{s}\bar{s}}$ states can be carried out and their evidences can also be searched for in experiments. Very recently, the LHCb Collaboration reported their results on the observations of tetraquark $cc\bar{c}\bar{c}$ ($T_{cc\bar{c}\bar{c}}$) states~\cite{Aaij:2020fnh}. A broad structure above the $J/\psi J/\psi$ threshold ranging from 6.2 to 6.8 GeV and a narrower structure $T_{cc\bar{c}\bar{c}}(6900)$ are observed with more than 5 $\sigma$ of significance level. There are also some vague structure around 7.2 GeV to be confirmed. These observations could be evidences for genuine $T_{cc\bar{c}\bar{c}}$ states~\cite{Lu:2020cns,Chen:2020xwe,Jin:2020jfc,Wang:2020ols,liu:2020eha}.

The observations of the
$T_{cc\bar{c}\bar{c}}$ states above the $J/\psi J/\psi$ threshold at LHCb may provide an important clue for the underlying dynamics for the ${cc\bar{c}\bar{c}}$ system. In particular, the narrowness of $T_{cc\bar{c}\bar{c}}(6900)$ suggests that the there should be more profound mechanism that ``slows down" the fall-apart decays of such a tetraquark system. Although this may be related to the properties of the static potential of heavy quark systems, more direct evidences are still needed to disentangle the dynamical features between the heavy and light flavor systems.
As an analogy of the $T_{cc\bar{c}\bar{c}}$ system, there might exist stable $T_{ss\bar{s}\bar{s}}$ states above
the $\phi\phi$ threshold, and can likely be observed in the di-$\phi$ mass spectrum. On the other hand, flavor mixings could be important for the light flavor systems and pure ${ss\bar{s}\bar{s}}$ states may not exist. To answer such questions, systematic calculations of the ${ss\bar{s}\bar{s}}$ system should be carried out. The BESIII experiments can provide a large data sample for the search of the $T_{ss\bar{s}\bar{s}}$ states in $J/\psi$ and $\psi(2S)$ decays.
In theory, although there have been some predictions of the $T_{ss\bar{s}\bar{s}}$ spectrum within the quark model
~\cite{Ebert:2008id,Lu:2019ira,tetra6} and QCD sum rules~\cite{Lu:2020cns,Chen:2020xwe,Jin:2020jfc,Wang:2020ols},
most of the studies focus on some special states in a diquark-antidiquark picture. About the status of
the tetraquark states, some recent review works can be referenced~\cite{Liu:2019zoy,Agaev:2020zad}.
In this study we intend to provide a systematical calculation of  the mass spectrum of the $1S$, $1P$ and $2S$-wave $T_{ss\bar{s}\bar{s}}$ states without the diquark-antidiquark approximation in a nonrelativistic potential quark model (NRPQM).

The NRPQM is based on the Hamiltonian proposed by the Cornell model~\cite{Eichten:1978tg},
which contains a linear confinement and a one-gluon-exchange (OGE) potential for
quark-quark and quark-antiquark interactions. With the NRPQM, we have successfully described
the $s\bar{s}$, $c\bar{c}$, and $b\bar{b}$ meson spectra~\cite{Li:2020xzs,Deng:2016stx,Deng:2016ktl}, and $sss$,
$ccc$ and $bbb$ baryon spectra~\cite{Liu:2019vtx,Liu:2019wdr}. Furthermore, we adopted
the NRPQM for the study of both $1S$ and $1P$-wave all-heavy tetraquark states with a
Gaussian expansion method~\cite{Liu:2019zuc,liu:2020eha}. In this work we
continue to extend this method to study the $T_{ss\bar{s}\bar{s}}$ spectrum by constructing the full tetraquark configurations without the
diquark-antidiquark approximation. With the parameters determined in our study of the $s\bar{s}$ spectrum~\cite{Li:2020xzs},
we obtain a relatively reliable
prediction of the mass spectrum for 36 $T_{ss\bar{s}\bar{s}}$ states, i.e., 4 $1S$-wave ground states, 20 $1P$-wave orbital excitations,
and 12 $2S$-wave radial excitations.

The paper is organized as follows: a brief introduction to the tetraquark spectrum
is given in Sec.~\ref{spectrum}. In Sec.~\ref{Results},
the numerical results and discussions are presented. A short summary is given in Sec.~\ref{Summary}.

\section{MASS SPECTRUM}\label{spectrum}

\subsection{Hamiltonian}

We adopt a NRPQM to calculate the mass spectrum of the $ss\bar{s}\bar{s}$ system. In this model the Hamiltonian is given by
\begin{equation}\label{Hamiltonian}
H=(\sum_{i=1}^4 m_i+T_i)-T_G+\sum_{i<j}V_{ij}(r_{ij}),
\end{equation}
where $m_i$ and $T_i$ stand for the constituent quark mass and kinetic energy of the $i$th quark, respectively; $T_G$ stands for the center-of-mass (c.m.) kinetic energy of the tetraquark system; $r_{ij}\equiv|\mathbf{r}_i-\mathbf{r}_j|$ is the distance between the $i$th and  $j$th quark; and $V_{ij}(r_{ij})$ stands for the effective potential between them.
In this work the $V_{ij}(r_{ij})$ adopts a widely used form
~\cite{Eichten:1978tg,Godfrey:1985xj,Swanson:2005,Godfrey:2015dia,Godfrey:2004ya,Lakhina:2006fy,Lu:2016bbk,Li:2010vx,Deng:2016stx,Deng:2016ktl}:
\begin{equation}\label{vij}
V_{ij}(r_{ij})=V_{ij}^{conf}(r_{ij})+V_{ij}^{sd}(r_{ij}),
\end{equation}
where the confinement potential adopts the standard form of the Cornell potential~\cite{Eichten:1978tg}, which includes
the spin-independent linear confinement potential $V_{ij}^{Lin}(r_{ij})\propto r_{ij}$ and Coulomb-like potential $V_{ij}^{Coul}(r_{ij})\propto 1/r_{ij}$:
\begin{equation}\label{vcon}
V_{ij}^{conf}(r_{ij})=-\frac{3}{16}({\vlab}_i\cdot{\vlab}_j)\left( b_{ij}r_{ij}-\frac{4}{3}\frac{\alpha_{ij}}{r_{ij}}+C_0 \right).
\end{equation}
The constant $C_0$ stands for the zero point energy. While the spin-dependent potential $V_{ij}^{sd}(r_{ij})$ is the sum of the spin-spin contact
hyperfine potential $V_{ij}^{SS}$, the spin-orbit potential $V_{ij}^{SS}$, and the tensor term $V_{ij}^{T}$:
\begin{equation}\label{voge OGE}
V^{sd}_{ij}(r_{ij})=V^{SS}_{ij}+V^{T}_{ij}+V^{LS}_{ij} \ ,
\end{equation}
%The spin-spin potential $V_{ij}^{SS}$, the spin-orbit potential $V_{ij}^{SS}$, and the tensor term $V_{ij}^{T}$ are given by
with
\begin{equation}\label{voge ss}
V^{SS}_{ij}=-\frac{\alpha_{ij}}{4}({\vlab}_i\cdot{\vlab}_j)\left\{\frac{\pi}{2}\cdot\frac{\sigma^3_{ij}e^{-\sigma^2_{ij}r_{ij}^2}}{\pi^{3/2}}\cdot\frac{16}{3m_im_j}(\mathbf{S}_i\cdot\mathbf{S}_j)\right\},
\end{equation}
\begin{equation}\label{voge ls}
\begin{split}
V^{LS}_{ij}=&-\frac{\alpha_{ij}}{16}\frac{{\vlab}_i\cdot{\vlab}_j}{r_{ij}^3} \bigg(\frac{1}{m_i^2}+\frac{1}{m_j^2}+\frac{4}{m_im_j}\bigg)\bigg\{\mathbf{L}_{ij}\cdot(\mathbf{S}_{i}+\mathbf{S}_{j})\bigg\}\\
&-\frac{\alpha_{ij}}{16}\frac{{\vlab}_i\cdot{\vlab}_j}{r_{ij}^3}\bigg(\frac{1}{m_i^2}-\frac{1}{m_j^2}\bigg)\bigg\{\mathbf{L}_{ij}\cdot(\mathbf{S}_{i}-\mathbf{S}_{j})\bigg\},
\end{split}
\end{equation}
\begin{equation}\label{voge ten}
V^{T}_{ij}=-\frac{\alpha_{ij}}{4}({\vlab}_i\cdot{\vlab}_j)\cdot\frac{1}{m_im_jr_{ij}^3}\Bigg\{\frac{3(\mathbf{S}_i\cdot \mathbf{r}_{ij})(\mathbf{S}_j\cdot \mathbf{r}_{ij})}{r_{ij}^2}-\mathbf{S}_i\cdot\mathbf{S}_j\Bigg\}.
\end{equation}
In the above equations, $\mathbf{S}_i$ stands for the spin of the $i$th quark,
and $\mathbf{L}_{ij}$ stands for the relative orbital angular momentum between the $i$th and $j$th quark.
If the interaction occurs between two quarks or antiquarks, the $\vlab_i\cdot\vlab_j$ operator is defined as
$\vlab_i\cdot\vlab_j\equiv\sum_{a=1}^8\lambda_i^a\lambda_j^a$, while if the interaction occurs between a quark and an antiquark, the $\vlab_i\cdot\vlab_j$ operator is defined as $\vlab_i\cdot\vlab_j\equiv\sum_{a=1}^8-\lambda_i^a\lambda_j^{a*}$, where $\lambda^{a*}$ is the complex conjugate of the Gell-Mann matrix $\lambda^a$. The parameters $b_{ij}$ and $\alpha_{ij}$ denote the strength of the confinement and strong coupling of the one-gluon-exchange potential, respectively.

The five parameters $m_s$, ${\alpha_{ss}}$, ${\sigma_{ss}}$, $b_{ss}$, and $C_0$ have been determined by fitting the mass spectrum of the strangeonium in our previous work~\cite{Li:2020xzs}. The quark model parameters adopted in this work are collected in the Table~\ref{parameters}.

\subsection{Configurations classified in the quark model}

To calculate the spectroscopy of a $qq\bar{q}\bar{q}$ ($q\in \{s,c,b\}$) system, first we construct the configurations
in the product space of flavor, color, spin, and spatial parts.

In the color space, there are two color-singlet bases $|6\bar{6}\rangle_c$ and $|\bar{3}3\rangle_c$, their wave functions are given by
\begin{equation}\label{colourf66}
\begin{split}
|6\bar{6}\rangle_c=&\frac{1}{2\sqrt{6}}\bigg[(rb+br)(\bar{b\mathstrut}\bar{r\mathstrut}+\bar{r\mathstrut}\bar{b\mathstrut})+(gr+rg)(\bar{g\mathstrut}\bar{r\mathstrut}+\bar{r\mathstrut}\bar{g\mathstrut})\\
&+(gb+bg)(\bar{b\mathstrut}\bar{g\mathstrut}+\bar{g\mathstrut}\bar{b\mathstrut})\\
&+2(rr)(\bar{r}\bar{r})+2(gg)(\bar{g}\bar{g})+2(bb)(\bar{b}\bar{b})\bigg],
\end{split}
\end{equation}
\begin{equation}\label{colourf33}
\begin{split}
|\bar{3}3\rangle_c=&\frac{1}{2\sqrt{3}}\bigg[(br-rb)(\bar{b\mathstrut}\bar{r\mathstrut}-\bar{r\mathstrut}\bar{b\mathstrut})-(rg-gr)(\bar{g\mathstrut}\bar{r\mathstrut}-\bar{r\mathstrut}\bar{g\mathstrut})\\
&+(bg-gb)(\bar{b\mathstrut}\bar{g\mathstrut}-\bar{g\mathstrut}\bar{b\mathstrut})\bigg].
\end{split}
\end{equation}

\begin{table}[htp]
\begin{center}
\caption{\label{parameters} Quark model parameters used in this work.}
\begin{tabular}{cccccccccccc}\hline\hline
%~~~~~~&  $m_u$~(GeV)           &~~~~~~~~~~~~~~~~~~~~~~~~~~~~~~~~~~~~~~0.313~~~~~~~~~~~~~\\
~~~~~~&  $m_s$~(GeV)           &~~~~~~~~~~~~~~~~~~~~~~~~~~~~~~~~~~~~~~0.60~~~~~~~~~~~~~\\
~~~~~~&  ${\alpha_{ss}}$       &~~~~~~~~~~~~~~~~~~~~~~~~~~~~~~~~~~~~~~0.77~~~~~~~~~~~~~\\
~~~~~~&  ${\sigma_{ss}}$~(GeV) &~~~~~~~~~~~~~~~~~~~~~~~~~~~~~~~~~~~~~~0.60~~~~~~~~~~~~~\\
~~~~~~&  ${b}$ ~(GeV $^2$)     &~~~~~~~~~~~~~~~~~~~~~~~~~~~~~~~~~~~~~~0.135~~~~~~~~~~~~~\\
~~~~~~&  ${C_0}$ ~(GeV)          &~~~~~~~~~~~~~~~~~~~~~~~~~~~~~~~~~~~~~~$-0.519$~~~~~~~~~~~~~\\
\hline\hline
\end{tabular}
\end{center}
\end{table}

In the spin space, there are six spin bases, which are denoted by $\chi_{S}^{S_{12}S_{34}}$.
Where $S_{12}$ stands for the spin quantum numbers for the diquark $(q_1q_2)$ (or
antidiquark $(\bar{q}_1\bar{q}_2)$), while $S_{34}$ stands for the spin quantum number
for the antidiquark $(\bar{q}_3\bar{q}_4)$ (or diquark $(q_3q_4)$ ). $S$ is the
total spin quantum number of the tetraquark $qq\bar{q}\bar{q}$ system. The spin
wave functions $\chi_{SS_z}^{S_{12}S_{34}}$ with a determined $S_z$ ($S_z$ stands for the third component of the total spin $\textbf{S}$) can be explicitly expressed as follows:
\begin{eqnarray}\label{spin000}
\chi_{00}^{00}&=&\frac{1}{2}(\uparrow\downarrow\uparrow\downarrow-\uparrow\downarrow\downarrow\uparrow-\downarrow\uparrow\uparrow\downarrow+\downarrow\uparrow\downarrow\uparrow),\\
\chi_{00}^{11}&=&\sqrt{\frac{1}{12}}(2\uparrow\uparrow\downarrow\downarrow-\uparrow\downarrow\uparrow\downarrow
-\uparrow\downarrow\downarrow\uparrow\nonumber \\
&&-\downarrow\uparrow\uparrow\downarrow-
\downarrow\uparrow\downarrow\uparrow+2\downarrow\downarrow\uparrow\uparrow),\\
\chi_{11}^{01}&=&\sqrt{\frac{1}{2}}(\uparrow\downarrow\uparrow\uparrow-\downarrow\uparrow\uparrow\uparrow),\\
\chi_{11}^{10}&=&\sqrt{\frac{1}{2}}(\uparrow\uparrow\uparrow\downarrow-\uparrow\uparrow\downarrow\uparrow),\\
\chi_{11}^{11}&=&\frac{1}{2}(\uparrow\uparrow\uparrow\downarrow+\uparrow\uparrow\downarrow\uparrow
-\uparrow\downarrow\uparrow\uparrow-\downarrow\uparrow\uparrow\uparrow),\\
\chi_{22}^{11}&=&\uparrow\uparrow\uparrow\uparrow.
\end{eqnarray}

In the spatial space, we define the relative Jacobi coordinates with the single-partial coordinates
$\mathbf{r_i}$ ($i=1,2,3,4$):
\begin{eqnarray}
\vxi_1&\equiv&\mathbf{r_1}-\mathbf{r_2},\\
\vxi_2&\equiv&\mathbf{r_3}-\mathbf{r_4},\\
\vxi_3&\equiv&\frac{m_1\mathbf{r_1}+m_2\mathbf{r_2}}{m_1+m_2}-\frac{m_3\mathbf{r_3}+m_4\mathbf{r_4}}{m_3+m_4},\\
\mathbf{R}&\equiv&\frac{m_1\mathbf{r_1}+m_2\mathbf{r_2}+m_3\mathbf{r_3}+m_4\mathbf{r_4}}{m_1+m_2+m_3+m_4}.
\end{eqnarray}
Note that $\vxi_1$ and $\vxi_2$ stand for the relative Jacobi coordinates between two quarks
$q_1$ and $q_2$ (or antiquarks $\bar{q}_1$ and $\bar{q}_2$), and two antiquarks $\bar{q}_3$ and $\bar{q}_4$ (or quarks $q_3$ and $q_4$), respectively. While $\vxi_3$ stands for the relative Jacobi coordinate between diquark $qq$ and anti-diquark $\bar{q}\bar{q}$.
Using the above Jacobi coordinates, it is easy to obtain basis functions that have well-defined symmetry
under permutations of the pairs $(12)$ and $(34)$~\cite{Vijande:2009kj}.

In the Jacobi coordinate system, the spatial wave function $\Psi_{NLM}(\vxi_1,\vxi_2,\vxi_3,\mathbf{R})$ for a $qq\bar{q}\bar{q}$ system with principal quantum number $N$ and orbital angular momentum quantum numbers $LM$ may be expressed as the linear
combination of $\Phi(\mathbf{R})\psi_{\alpha_1}(\vxi_1)\psi_{\alpha_2}(\vxi_2)\psi_{\alpha_3}(\vxi_3)$:
\begin{eqnarray}\label{cf1}
\Psi_{NLM}(\vxi_1,\vxi_2,\vxi_3,\mathbf{R})\ \ \ \ \ \ \ \ \ \ \ \ \ \ \ \ \ \ \ \ \ \ \ \ \ \ \ \ \ \ \ \ \ \ \ \ \ \ \ \ \ \ \ \ \ \ \ \ \ \ \ \ \ \nonumber \\
=\sum_{\alpha_1,\alpha_2,\alpha_3}C_{\alpha_1,\alpha_2,\alpha_3}[\Phi(\mathbf{R})\psi_{\alpha_1}(\vxi_1)\psi_{\alpha_2}(\vxi_2)\psi_{\alpha_3}(\vxi_3)]_{NLM},
\end{eqnarray}
where $C_{\alpha_1,\alpha_2,\alpha_3}$ stands for the combination coefficients, $\Phi(\mathbf{R})$ is the center-of-mass (c.m.) motion wave function.
In the quantum number set $\alpha_i\equiv \{n_{\xi_i},l_{\xi_i}, m_{\xi_i}\}$,
$n_{\xi_i}$ is the principal quantum number, $l_{\xi_i}$ is the angular momentum,
and $m_{\xi_i}$ is its third component projection.
The wave functions $\psi_{\alpha_i}(\vxi_i)$, which account for the relative motions, can be written as
\begin{equation}\label{cf2}
\psi_{\alpha_i}(\vxi_i)=R_{n_{\xi_i}l_{\xi_i}}(\xi_i) Y_{l_{\xi_i}m_{\xi_i}}(\hat{\vxi_i}),
\end{equation}
where $Y_{l_{\xi_i}m_{\xi_i}}(\hat{\vxi_i})$ is the spherical harmonic function, and
$R_{n_{\xi_i}l_{\xi_i}}(\xi_i)$ is the radial part. It is seen that for an excited state, there
are three spatial excitation modes corresponding to
three independent internal wave functions $\psi_{\alpha_i}(\vxi_i)$ ($i=1,2,3$), which are denoted as
$\xi_1$, $\xi_2$, and $\xi_3$, respectively, in the present work.
One point should be emphasized that considering the fact that the $ss\bar{s}\bar{s}$ system is composed of equal mass constituent quarks and antiquarks, we adopt a single set of Jacobi coordinates in this study as an approximation.
In fact, the four-body wave function describing a scalar $ss\bar{s}\bar{s}$ state contains
a small contribution of internal angular momentum.
This contribution is neglected in our calculations. To precisely treat an $N$-body system, one can
involve several different sets of Jacobi coordinates as those done in Refs.~\cite{Hiyama:2003cu,Hiyama:2018ivm,Meng:2020knc,Hiyama:2005cf,Hiyama:2018ukv,Meng:2019fan};
or adopt a single set of Jacobi coordinates $X=(\vxi_i, \vxi_2,..., \vxi_{N-1})$ with
non diagonal Gaussians $e^{-XAX^T}$ as those done in Refs.~\cite{Varga:1996zz,Varga:1996jr,Varga:1997xga,Brink:1998as},
where $A$ is a symmetric matrix.

Taking into account the Pauli principle and color confinement for the four-quark system $qq\bar{q}\bar{q}$, we
have 4 configurations for $1S$-wave ground states, 20 configurations for the $1P$-wave orbital excitations,
and 12 configurations for the $2S$-wave radial excitations. The spin-parity quantum numbers, notations, and wave functions
for these configurations are presented in Table~\ref{states}. With the wave functions for all the configurations,
the mass matrix elements of the Hamiltonian can be worked out.

To work out the matrix elements in the coordinate space, we expand the radial part $R_{n_{\xi_i}l_{\xi_i}}(\xi_i)$ with a series of harmonic oscillator functions~\cite{Liu:2019vtx,liu:2020eha}:
\begin{equation}\label{spatial function1}
R_{n_{\xi_i}l_{\xi_i}}(\xi_i)=\sum_{\ell=1}^n\mathcal{C}_{\xi_i\ell}~\phi_{n_{\xi_i} l_{\xi_i}}(\omega_{\xi_i\ell},\xi_i),
\end{equation}
with
\begin{eqnarray}\label{spatial function2}
\phi_{n_{\xi_i} l_{\xi_i}}(\omega_{\xi_i\ell},\xi_i)&=\left(\mu_{\xi_i}\omega_{\xi_i\ell}\right)^{\frac{3}{4}}\Bigg[\frac{2^{l_\xi+2-n_\xi}
(2l_{\xi_i}+2n_{\xi_i}+1)!!}{\sqrt{\pi}n_{\xi_i}![(2l_{\xi_i}+1)!!]^2}\Bigg]^{\frac{1}{2}}\left(\sqrt{\mu_{\xi_i}\omega_{{\xi_i}\ell}}\xi_i\right)^{l_{\xi_i}}\nonumber\\
&\times e^{-\frac{1}{2}\mu_{\xi_i}\omega_{\xi_i\ell}{\xi_i}^2}F\left(-n_{\xi_i},l_{\xi_i}+\frac{3}{2},\mu_{\xi_i}\omega_{\xi_i\ell}{\xi_i}^2\right),
\end{eqnarray}
where $F\left(-n_{\xi_i},l_{\xi_i}+\frac{3}{2},\mu_{\xi_i}\omega_{\xi_i\ell}{\xi_i}^2\right)$ is the confluent hypergeometric function.
It should be pointed out that if there are no radial excitations, the expansion method with harmonic oscillator wave
functions are just the same as the Gaussian expansion method adopted in the literature~\cite{Hiyama:2003cu,Hiyama:2018ivm}.

\begin{figure}
\centering \epsfxsize=9 cm \epsfbox{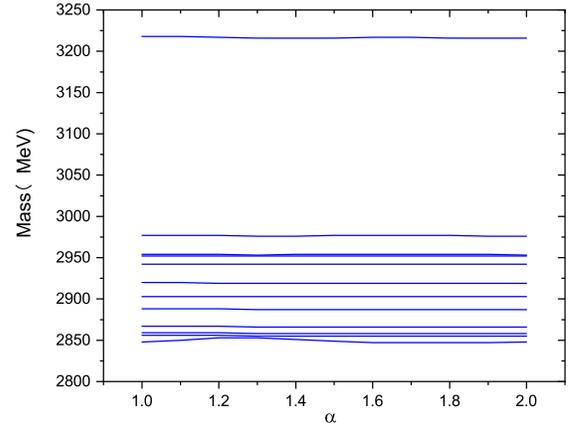} \vspace{-0.9 cm} \caption{Predicted masses of 12 $2S$-wave $T_{ss\bar{s}\bar{s}}$ configurations as a function of the scaling factor $\alpha$.}\label{figsa}
\end{figure}

For an $ss\bar{s}\bar{s}$ system, if we ensure that the spatial wave function with Jacobi coordinates
can transform into the single particle coordinate system,
the harmonic oscillator frequencies $\omega_{\xi_i\ell}$ ($i=1,2,3$) can be related to the harmonic oscillator stiffness factor $K_{\ell}$ with $\omega_{\xi_1\ell}=\sqrt{2K_\ell/\mu_{\xi_1}}$, $\omega_{\xi_2\ell}=\sqrt{2K_\ell/\mu_{\xi_2}}$, and $\omega_{\xi_3\ell}=\sqrt{4K_\ell/\mu_{\xi_3}}$.
Considering the reduced masses $\mu_{\xi_1}=\mu_{\xi_2}=m_s/2$, $\mu_{\xi_3}=m_s$ for $T_{(ss\bar{s}\bar{s})}$,
one has $\omega_{\xi_1\ell}=\omega_{\xi_2\ell}=\omega_{\xi_3\ell}=\sqrt{4K_\ell/m_s}$.
It indicates that the harmonic oscillator frequencies $\omega_{\xi_i\ell}$ for $T_{(ss\bar{s}\bar{s})}$ are not independent.
According to the relation $\omega_{\xi_1\ell}=\omega_{\xi_2\ell}=\omega_{\xi_3\ell}=\omega_{\ell}$,
the expansion of $\prod_{i=1}^3 R_{n_{\xi_i} l_{\xi_i}}({\xi_i})$ can be simplified as
\begin{widetext}
\begin{equation}
\begin{split}
\prod_{i=1}^3 R_{n_{\xi_i} l_{\xi_i}}({\xi_i})
=&\sum_{\ell}^n\sum_{\ell'}^n\sum_{\ell''}^n \mathcal{C}_{{\xi_1}\ell} \mathcal{C}_{{\xi_2}\ell'} \mathcal{C}_{{\xi_3}\ell''}
~\phi_{n_{\xi_1} l_{\xi_1}}(\omega_{{\xi_1}\ell},{\xi_1})
~\phi_{n_{\xi_2} l_{\xi_2}}(\omega_{{\xi_2}\ell'},{\xi_2})
~\phi_{n_{\xi_3} l_{\xi_3}}(\omega_{{\xi_3}\ell''},{\xi_3})
~\delta_{\ell\ell'}\delta_{\ell\ell''}\\
=&\sum_{\ell}^n \mathcal{C}_{\ell}
~\phi_{n_{\xi_1} l_{\xi_1}}(\omega_{\ell},{\xi_1})
~\phi_{n_{\xi_2} l_{\xi_2}}(\omega_{\ell},{\xi_2})
~\phi_{n_{\xi_3} l_{\xi_3}}(\omega_{\ell},{\xi_3}).
\end{split}
\end{equation}
\end{widetext}
Then we introduce oscillator length parameters $d_{\ell}$ that can be related to the
harmonic oscillator frequencies $\omega_{\ell}$ with $1/d^2_{\ell}=m_s\omega_{\ell}$.
Following the method of Refs.~\cite{Hiyama:2003cu,Hiyama:2018ivm}, we let the $d_\ell$ parameters form a geometric progression
\begin{equation}\label{geometric progression}
d_\ell=d_1a^{\ell-1}\ \ \ (\ell=1,...,n),
\end{equation}
where $n$ is the number of harmonic oscillator functions, and $a$ is the ratio coefficient. There are three parameters $\{d_1,d_n,n\}$ to be determined through the variation method. It is found that with the parameter set \{0.085 fm, 3.399 fm, 15\} for the $ss\bar{s}\bar{s}$ system, we can obtain stable solutions. The numerical results should be independent of the parameter $d_1$. To confirm this point,
as done in the literature~\cite{Hiyama:2005cf,Hiyama:2018ukv,Meng:2019fan} we scale the parameter $d_1$ of the basis functions as $d_1\to \alpha d_1$. The mass of a $T_{ss\bar{s}\bar{s}}$ state should be stable at a resonance
energy insensitive to the scaling parameter $\alpha$. As an example,
we plot the masses of 12 $2S$-wave $T_{ss\bar{s}\bar{s}}$ configurations as a
function of the scaling factor $\alpha$ in Fig.~\ref{figsa}.
It is found that the numerical results are nearly independent of the scaling factor $\alpha$.
The stabilization of other states predicted in this work has also been examined by the same method.

With the mass matrix elements ready for each configuration,
the mass of the tetraquark configuration and its spacial wave function can be determined by solving a generalized eigenvalue problem.
The details can be found in our previous works~\cite{Liu:2019vtx,Liu:2019zuc}. Finally, the physical states can be obtained by diagonalizing the mass matrix of different configurations with the same $J^{PC}$ numbers.

\begin{table*}[htp]
\begin{center}
\caption{\label{states} Configurations for the tetraquark $qq\bar{q}\bar{q}$ system up to the $2S$-wave states.}
\begin{tabular}{cclc}
\hline\hline
$J^{P(C)}$  & Configuration  & \multicolumn{2}{c}{\underline{~~~~~~~~~~~~~~~~~~~~~~~~~~~~~~~~~~~~~~~~~~~~~~~~~Wave Function~~~~~~~~~~~~~~~~~~~~~~~~~~~~~~~~~~~~~~~~~~~~~~~~~~~}}\tabularnewline
\hline\hline
$J^{PC}$=$0^{++}$  & $1^{1}S_{0^{++}(6\bar{6})_{c}}$ & $\psi_{000}^{1S}\chi_{00}^{00}$ & $|6\bar{6}\rangle^{c}$\tabularnewline
$J^{PC}$=$0^{++}$  & $1^{1}S_{0^{++}(\bar{3}3)_{c}}$ & $\psi_{000}^{1S}\chi_{00}^{11}$  & $|\bar{3}3\rangle^{c}$\tabularnewline
$J^{PC}$=$1^{+-}$  & $1^{3}S_{1^{+-}(\bar{3}3)_{c}}$ & $\psi_{000}^{1S}\chi_{11}^{11}$  & $|\bar{3}3\rangle^{c}$\tabularnewline
$J^{PC}$=$2^{++}$  & $1^{5}S_{2^{++}(\bar{3}3)_{c}}$ & $\psi_{000}^{1S}\chi_{22}^{11}$  & $|\bar{3}3\rangle^{c}$\tabularnewline
\hline
$J^{PC}$=$0^{--}$  & $1^{3}P_{0^{--}(6\bar{6})_{c}\left(\xi_{1},\xi_{2}\right)}$ & $\sqrt{\frac{1}{6}}\left(\psi_{011}^{\xi_1}\chi_{1-1}^{10}-\psi_{010}^{\xi_1}\chi_{10}^{10}+\psi_{01-1}^{\xi_1}\chi_{11}^{10}-\psi_{011}^{\xi_2}\chi_{1-1}^{01}+\psi_{010}^{\xi_2}\chi_{10}^{01}-\psi_{01-1}^{\xi_2}\chi_{11}^{01}\right)$  & $|6\bar{6}\rangle^{c}$\tabularnewline
$J^{PC}$=$0^{--}$  & $1^{3}P_{0^{--}(\bar{3}3)_{c}\left(\xi_{1},\xi_{2}\right)}$ & $\sqrt{\frac{1}{6}}\left(\psi_{011}^{\xi_1}\chi_{1-1}^{01}-\psi_{010}^{\xi_1}\chi_{10}^{01}+\psi_{01-1}^{\xi_1}\chi_{11}^{01}-\psi_{011}^{\xi_2}\chi_{1-1}^{10}+\psi_{010}^{\xi_2}\chi_{10}^{10}-\psi_{01-1}^{\xi_2}\chi_{11}^{10}\right)$  & $|\bar{3}3\rangle^{c}$\tabularnewline
$J^{PC}$=$0^{-+}$  & $1^{3}P_{0^{-+}(6\bar{6})_{c}\left(\xi_{1},\xi_{2}\right)}$ & $\sqrt{\frac{1}{6}}\left(\psi_{011}^{\xi_1}\chi_{1-1}^{10}-\psi_{010}^{\xi_1}\chi_{10}^{10}+\psi_{01-1}^{\xi_1}\chi_{11}^{10}+\psi_{011}^{\xi_2}\chi_{1-1}^{01}-\psi_{010}^{\xi_2}\chi_{10}^{01}+\psi_{01-1}^{\xi_2}\chi_{11}^{01}\right)$  & $|6\bar{6}\rangle^{c}$\tabularnewline
$J^{PC}$=$0^{-+}$  & $1^{3}P_{0^{-+}(\bar{3}3)_{c}\left(\xi_{1},\xi_{2}\right)}$  & $\sqrt{\frac{1}{6}}\left(\psi_{011}^{\xi_1}\chi_{1-1}^{01}-\psi_{010}^{\xi_1}\chi_{10}^{01}+\psi_{01-1}^{\xi_1}\chi_{11}^{01}+\psi_{011}^{\xi_2}\chi_{1-1}^{10}-\psi_{010}^{\xi_2}\chi_{10}^{10}+\psi_{01-1}^{\xi_2}\chi_{11}^{10}\right)$  & $|\bar{3}3\rangle^{c}$\tabularnewline
$J^{PC}$=$0^{-+}$  & $1^{3}P_{0^{-+}(\bar{3}3)_{c}\left(\xi_{3}\right)}$  & $\sqrt{\frac{1}{3}}\left(\psi_{011}^{\xi_3}\chi_{1-1}^{11}-\psi_{010}^{\xi_3}\chi_{10}^{11}+\psi_{01-1}^{\xi_3}\chi_{11}^{11}\right)$  & $|\bar{3}3\rangle^{c}$\tabularnewline
$J^{PC}$=$1^{--}$  & $1^{3}P_{1^{--}(6\bar{6})_{c}\left(\xi_{1},\xi_{2}\right)}$ & $\frac{1}{2}\left(\psi_{011}^{\xi_1}\chi_{10}^{10}-\psi_{010}^{\xi_1}\chi_{11}^{10}-\psi_{011}^{\xi_2}\chi_{10}^{01}+\psi_{010}^{\xi_2}\chi_{11}^{01}\right)$  & $|6\bar{6}\rangle^{c}$\tabularnewline
$J^{PC}$=$1^{--}$  & $1^{3}P_{1^{--}(\bar{3}3)_{c}\left(\xi_{1},\xi_{2}\right)}$  & $\frac{1}{2}\left(\psi_{011}^{\xi_1}\chi_{10}^{01}-\psi_{010}^{\xi_1}\chi_{11}^{01}-\psi_{011}^{\xi_2}\chi_{10}^{10}+\psi_{010}^{\xi_2}\chi_{11}^{10}\right)$  & $|\bar{3}3\rangle^{c}$\tabularnewline
$J^{PC}$=$1^{--}$  & $1^{5}P_{1^{--}(\bar{3}3)_{c}\left(\xi_{3}\right)}$  & $\sqrt{\frac{1}{10}}\psi_{011}^{\xi_3}\chi_{20}^{11}-\sqrt{\frac{3}{10}}\psi_{010}^{\xi_3}\chi_{21}^{11}+\sqrt{\frac{3}{5}}\psi_{01-1}^{\xi_3}\chi_{22}^{11}$  & $|\bar{3}3\rangle^{c}$\tabularnewline
$J^{PC}$=$1^{--}$  & $1^{1}P_{1^{--}(\bar{3}3)_{c}\left(\xi_{3}\right)}$  & $\psi_{011}^{\xi_3}\chi_{00}^{11}$ & $|\bar{3}3\rangle^{c}$\tabularnewline
$J^{PC}$=$1^{--}$  & $1^{1}P_{1^{--}(6\bar{6})_{c}\left(\xi_{3}\right)}$ & $\psi_{011}^{\xi_3}\chi_{00}^{00}$ & $|6\bar{6}\rangle^{c}$\tabularnewline
$J^{PC}$=$1^{-+}$  & $1^{3}P_{1^{-+}(6\bar{6})_{c}\left(\xi_{1},\xi_{2}\right)}$ & $\frac{1}{2}\left(\psi_{011}^{\xi_1}\chi_{10}^{10}-\psi_{010}^{\xi_1}\chi_{11}^{10}+\psi_{011}^{\xi_2}\chi_{10}^{01}-\psi_{010}^{\xi_2}\chi_{11}^{01}\right)$  & $|6\bar{6}\rangle^{c}$\tabularnewline
$J^{PC}$=$1^{-+}$  & $1^{3}P_{1^{-+}(\bar{3}3)_{c}\left(\xi_{1},\xi_{2}\right)}$  & $\frac{1}{2}\left(\psi_{011}^{\xi_1}\chi_{10}^{01}-\psi_{010}^{\xi_1}\chi_{11}^{01}+\psi_{011}^{\xi_2}\chi_{10}^{10}-\psi_{010}^{\xi_2}\chi_{11}^{10}\right)$  & $|\bar{3}3\rangle^{c}$\tabularnewline
$J^{PC}$=$1^{-+}$  & $1^{3}P_{1^{-+}(\bar{3}3)_{c}\left(\xi_{3}\right)}$  & $\sqrt{\frac{1}{2}}\left(\psi_{011}^{\xi_3}\chi_{10}^{11}-\psi_{010}^{\xi_3}\chi_{11}^{11}\right)$  & $|\bar{3}3\rangle^{c}$\tabularnewline
$J^{PC}$=$2^{--}$  & $1^{3}P_{2^{--}(6\bar{6})_{c}\left(\xi_{1},\xi_{2}\right)}$ & $\sqrt{\frac{1}{2}}\left(\psi_{011}^{\xi_1}\chi_{11}^{10}-\psi_{011}^{\xi_2}\chi_{11}^{01}\right)$  & $|6\bar{6}\rangle^{c}$\tabularnewline
$J^{PC}$=$2^{--}$  & $1^{3}P_{2^{--}(\bar{3}3)_{c}\left(\xi_{1},\xi_{2}\right)}$  & $\sqrt{\frac{1}{2}}\left(\psi_{011}^{\xi_1}\chi_{11}^{01}-\psi_{011}^{\xi_2}\chi_{11}^{10}\right)$  & $|\bar{3}3\rangle^{c}$\tabularnewline
$J^{PC}$=$2^{--}$  & $1^{5}P_{2^{--}(\bar{3}3)_{c}\left(\xi_{3}\right)}$  & $\sqrt{\frac{1}{3}}\psi_{011}^{\xi_3}\chi_{21}^{11}-\sqrt{\frac{2}{3}}\psi_{010}^{\xi_3}\chi_{22}^{11}$  & $|\bar{3}3\rangle^{c}$\tabularnewline
$J^{PC}$=$2^{-+}$  & $1^{3}P_{2^{-+}(6\bar{6})_{c}\left(\xi_{1},\xi_{2}\right)}$ & $\sqrt{\frac{1}{2}}\left(\psi_{011}^{\xi_1}\chi_{11}^{10}+\psi_{011}^{\xi_2}\chi_{11}^{01}\right)$  & $|6\bar{6}\rangle^{c}$\tabularnewline
$J^{PC}$=$2^{-+}$  & $1^{3}P_{2^{-+}(\bar{3}3)_{c}\left(\xi_{1},\xi_{2}\right)}$  & $\sqrt{\frac{1}{2}}\left(\psi_{011}^{\xi_1}\chi_{11}^{01}+\psi_{011}^{\xi_2}\chi_{11}^{10}\right)$  & $|\bar{3}3\rangle^{c}$\tabularnewline
$J^{PC}$=$2^{-+}$  & $1^{3}P_{2^{-+}(\bar{3}3)_{c}\left(\xi_{3}\right)}$  & $\psi_{011}^{\xi_3}\chi_{11}^{11}$  & $|\bar{3}3\rangle^{c}$\tabularnewline
$J^{PC}$=$3^{--}$  & $1^{5}P_{3^{--}(\bar{3}3)_{c}\left(\xi_{3}\right)}$  & $\psi_{011}^{\xi_3}\chi_{22}^{11}$ & $|\bar{3}3\rangle^{c}$\tabularnewline
\hline
$J^{PC}$=$0^{+-}$  & $2^{1}S_{0^{+-}(6\bar{6})_{c}\left(\xi_{1},\xi_{2}\right)}$ & $\sqrt{\frac{1}{2}}\left(\psi_{100}^{\xi_1}\chi_{00}^{00}-\psi_{100}^{\xi_2}\chi_{00}^{00}\right)$  & $|6\bar{6}\rangle^{c}$\tabularnewline
$J^{PC}$=$0^{+-}$  & $2^{1}S_{0^{+-}(\bar{3}3)_{c}\left(\xi_{1},\xi_{2}\right)}$ & $\sqrt{\frac{1}{2}}\left(\psi_{100}^{\xi_1}\chi_{00}^{11}-\psi_{100}^{\xi_2}\chi_{00}^{11}\right)$  & $|\bar{3}3\rangle^{c}$\tabularnewline
$J^{PC}$=$0^{++}$  & $2^{1}S_{0^{++}(6\bar{6})_{c}\left(\xi_{1},\xi_{2}\right)}$ & $\sqrt{\frac{1}{2}}\left(\psi_{100}^{\xi_1}\chi_{00}^{00}+\psi_{100}^{\xi_2}\chi_{00}^{00}\right)$ & $|6\bar{6}\rangle^{c}$\tabularnewline
$J^{PC}$=$0^{++}$  & $2^{1}S_{0^{++}(\bar{3}3)_{c}\left(\xi_{1},\xi_{2}\right)}$  & $\sqrt{\frac{1}{2}}\left(\psi_{100}^{\xi_1}\chi_{00}^{11}+\psi_{100}^{\xi_2}\chi_{00}^{11}\right)$  & $|\bar{3}3\rangle^{c}$\tabularnewline
$J^{PC}$=$0^{++}$  & $2^{1}S_{0^{++}(6\bar{6})_{c}\left(\xi_{3}\right)}$  & $\psi_{100}^{\xi_3}\chi_{00}^{00}$  & $|6\bar{6}\rangle^{c}$\tabularnewline
$J^{PC}$=$0^{++}$  & $2^{1}S_{0^{++}(\bar{3}3)_{c}\left(\xi_{3}\right)}$  & $\psi_{100}^{\xi_3}\chi_{00}^{11}$  & $|\bar{3}3\rangle^{c}$\tabularnewline
$J^{PC}$=$1^{+-}$  & $2^{3}S_{1^{+-}(\bar{3}3)_{c}\left(\xi_{1},\xi_{2}\right)}$ & $\sqrt{\frac{1}{2}}\left(\psi_{100}^{\xi_1}\chi_{11}^{11}+\psi_{100}^{\xi_2}\chi_{11}^{11}\right)$  & $|\bar{3}3\rangle^{c}$\tabularnewline
$J^{PC}$=$1^{+-}$  & $2^{3}S_{1^{+-}(\bar{3}3)_{c}\left(\xi_{3}\right)}$  & $\psi_{100}^{\xi_3}\chi_{11}^{11}$  & $|\bar{3}3\rangle^{c}$\tabularnewline
$J^{PC}$=$1^{++}$  & $\begin{array}{c}
2^{3}S_{1^{++}(\bar{3}3)_{c}\left(\xi_{1},\xi_{2}\right)}\end{array}$  & $\sqrt{\frac{1}{2}}\left(\psi_{100}^{\xi_1}\chi_{11}^{11}-\psi_{100}^{\xi_2}\chi_{11}^{11}\right)$  & $|\bar{3}3\rangle^{c}$\tabularnewline
$J^{PC}$=$2^{+-}$  & $\begin{array}{c}
2^{5}S_{2^{+-}(\bar{3}3)_{c}\left(\xi_{1},\xi_{2}\right)}\end{array}$  & $\sqrt{\frac{1}{2}}\left(\psi_{100}^{\xi_1}\chi_{22}^{11}-\psi_{100}^{\xi_2}\chi_{22}^{11}\right)$  & $|\bar{3}3\rangle^{c}$\tabularnewline
$J^{PC}$=$2^{++}$  & $2^{5}S_{2^{++}(\bar{3}3)_{c}\left(\xi_{1},\xi_{2}\right)}$  & $\sqrt{\frac{1}{2}}\left(\psi_{100}^{\xi_1}\chi_{22}^{11}+\psi_{100}^{\xi_2}\chi_{22}^{11}\right)$  & $|\bar{3}3\rangle^{c}$\tabularnewline
$J^{PC}$=$2^{++}$  & $2^{5}S_{2^{++}(\bar{3}3)_{c}\left(\xi_{3}\right)}$  & $\psi_{100}^{\xi_3}\chi_{22}^{11}$  & $|\bar{3}3\rangle^{c}$\tabularnewline
\hline\hline
\end{tabular}
\end{center}
\end{table*}

\begin{table*}[htp]
\begin{center}
\caption{\label{mass of ssss} Predicted mass spectrum for the $ss\bar{s}\bar{s}$ system with the HOEM.}
\begin{tabular}{ccccc}
\hline
$J^{P(C)}$ & Configuration & $\langle H\rangle$ (MeV) & Mass (MeV) & Eigenvector\tabularnewline
\hline
$0^{++}$ & $\begin{array}{l}
1^{1}S_{0^{++}(6\bar{6})_{c}}\\
1^{1}S_{0^{++}(\bar{3}3)_{c}}
\end{array}$ & $\left(\begin{array}{cc}
2365 & -105\\
-105 & 2293
\end{array}\right)$ & $\left(\begin{array}{c}
2218\\
2440
\end{array}\right)$ & $\left(\begin{array}{c}
\left(\begin{array}{cc}
-0.58 & -0.81\end{array}\right)\\
\left(\begin{array}{cc}
-0.81 & 0.58\end{array}\right)
\end{array}\right)$\tabularnewline
$1^{+-}$ & $\begin{array}{l}
1^{3}S_{1^{+-}(\bar{3}3)_{c}}\end{array}$ & $\left(2323\right)$ & $2323$ & $1$\tabularnewline
$2^{++}$ & $\begin{array}{l}
1^{5}S_{2^{++}(\bar{3}3)_{c}}\end{array}$ & $\left(2378\right)$ & $2378$ & $1$\tabularnewline
\hline
$0^{--}$ & $\begin{array}{l}
^{3}P_{0^{--}(6\bar{6})_{c}\left(\xi_{1},\xi_{2}\right)}\\
^{3}P_{0^{--}(\bar{3}3)_{c}\left(\xi_{1},\xi_{2}\right)}
\end{array}$ & $\left(\begin{array}{cc}
2635 & 154\\
154 & 2694
\end{array}\right)$ & $\left(\begin{array}{c}
2507\\
2821
\end{array}\right)$ & $\left(\begin{array}{c}
\left(\begin{array}{cc}
-0.77 & 0.64\end{array}\right)\\
\left(\begin{array}{cc}
0.64 & 0.77\end{array}\right)
\end{array}\right)$\tabularnewline
$0^{-+}$ & $\begin{array}{l}
^{3}P_{0^{-+}(6\bar{6})_{c}\left(\xi_{1},\xi_{2}\right)}\\
^{3}P_{0^{-+}(\bar{3}3)_{c}\left(\xi_{1},\xi_{2}\right)}\\
^{3}P_{0^{-+}(\bar{3}3)_{c}\left(\xi_{3}\right)}
\end{array}$ & $\left(\begin{array}{ccc}
2616 & -35 & -111\\
-35 & 2685 & 56\\
-111 & 56 & 2576
\end{array}\right)$ & $\left(\begin{array}{c}
2481\\
2635\\
2761
\end{array}\right)$ & $\left(\begin{array}{c}
\left(\begin{array}{ccc}
0.61 & -0.11 & 0.78\end{array}\right)\\
\left(\begin{array}{ccc}
-0.56 & -0.76 & 0.34\end{array}\right)\\
\left(\begin{array}{ccc}
0.56 & -0.64 & -0.53\end{array}\right)
\end{array}\right)$\tabularnewline
$1^{--}$ & $\begin{array}{l}
^{3}P_{1^{--}(6\bar{6})_{c}\left(\xi_{1},\xi_{2}\right)}\\
^{3}P_{1^{--}(\bar{3}3)_{c}\left(\xi_{1},\xi_{2}\right)}\\
^{5}P_{1^{--}(\bar{3}3)_{c}\left(\xi_{3}\right)}\\
^{1}P_{1^{--}(\bar{3}3)_{c}\left(\xi_{3}\right)}\\
^{1}P_{1^{--}(6\bar{6})_{c}\left(\xi_{3}\right)}
\end{array}$ & $\left(\begin{array}{ccccc}
2585 & -154 & -89 & -46 & 90\\
-154 & 2694 & 42 & 22 & -76\\
-89 & 42 & 2584 & -8 & 29\\
-46 & 22 & -8 & 2636 & -51\\
90 & -76 & 29 & -51 & 2889
\end{array}\right)$ & $\left(\begin{array}{c}
2445\\
2567\\
2627\\
2766\\
2984
\end{array}\right)$ & $\left(\begin{array}{c}
\left(\begin{array}{ccccc}
-0.80 & -0.38 & -0.43 & -0.14 & 0.11\end{array}\right)\\
\left(\begin{array}{ccccc}
0.18 & 0.57 & -0.78 & -0.05 & 0.15\end{array}\right)\\
\left(\begin{array}{ccccc}
0.03 & 0.11 & 0.11 & -0.97 & -0.18\end{array}\right)\\
\left(\begin{array}{ccccc}
-0.42 & 0.57 & 0.43 & 0.00 & 0.56\end{array}\right)\\
\left(\begin{array}{ccccc}
0.38 & -0.43 & -0.07 & -0.19 & 0.79\end{array}\right)
\end{array}\right)$\tabularnewline
$1^{-+}$ & $\begin{array}{l}
^{3}P_{1^{-+}(6\bar{6})_{c}\left(\xi_{1},\xi_{2}\right)}\\
^{3}P_{1^{-+}(\bar{3}3)_{c}\left(\xi_{1},\xi_{2}\right)}\\
^{3}P_{1^{-+}(\bar{3}3)_{c}\left(\xi_{3}\right)}
\end{array}$ & $\left(\begin{array}{ccc}
2628 & 95 & 25\\
95 & 2712 & 12\\
25 & 12 & 2633
\end{array}\right)$ & $\left(\begin{array}{c}
2564\\
2632\\
2778
\end{array}\right)$ & $\left(\begin{array}{c}
\left(\begin{array}{ccc}
0.83 & -0.51 & -0.21\end{array}\right)\\
\left(\begin{array}{ccc}
-0.09 & 0.25 & -0.96\end{array}\right)\\
\left(\begin{array}{ccc}
0.55 & 0.82 & 0.17\end{array}\right)
\end{array}\right)$\tabularnewline
$2^{--}$ & $\begin{array}{l}
^{3}P_{2^{--}(6\bar{6})_{c}\left(\xi_{1},\xi_{2}\right)}\\
^{3}P_{2^{--}(\bar{3}3)_{c}\left(\xi_{1},\xi_{2}\right)}\\
^{5}P_{2^{--}(\bar{3}3)_{c}\left(\xi_{3}\right)}
\end{array}$ & $\left(\begin{array}{ccc}
2620 & -217 & -50\\
-217 & 2725 & 24\\
-50 & 24 & 2665
\end{array}\right)$ & $\left(\begin{array}{c}
2446\\
2657\\
2907
\end{array}\right)$ & $\left(\begin{array}{c}
\left(\begin{array}{ccc}
0.79 & 0.60 & 0.12\end{array}\right)\\
\left(\begin{array}{ccc}
-0.03 & 0.23 & -0.97\end{array}\right)\\
\left(\begin{array}{ccc}
-0.61 & 0.76 & 0.20\end{array}\right)
\end{array}\right)$\tabularnewline
$2^{-+}$ & $\begin{array}{l}
^{3}P_{2^{-+}(6\bar{6})_{c}\left(\xi_{1},\xi_{2}\right)}\\
^{3}P_{2^{-+}(\bar{3}3)_{c}\left(\xi_{1},\xi_{2}\right)}\\
^{3}P_{2^{-+}(\bar{3}3)_{c}\left(\xi_{3}\right)}
\end{array}$ & $\left(\begin{array}{ccc}
2638 & 138 & -33\\
138 & 2733 & -16\\
-33 & -16 & 2673
\end{array}\right)$ & $\left(\begin{array}{c}
2537\\
2669\\
2837
\end{array}\right)$ & $\left(\begin{array}{c}
\left(\begin{array}{ccc}
0.82 & -0.56 & 0.13\end{array}\right)\\
\left(\begin{array}{ccc}
0.00 & 0.23 & 0.97\end{array}\right)\\
\left(\begin{array}{ccc}
-0.58 & -0.79 & 0.19\end{array}\right)
\end{array}\right)$\tabularnewline
$3^{--}$ & $\begin{array}{c}
^{5}P_{3^{--}(\bar{3}3)_{c}\left(\xi_{3}\right)}\end{array}$ & $\left(\begin{array}{c}
2719\end{array}\right)$ & $2719$ & $1$\tabularnewline
\hline
$0^{+-}$ & $\begin{array}{l}
2^{1}S_{0^{+-}(6\bar{6})_{c}\left(\xi_{1},\xi_{2}\right)}\\
2^{1}S_{0^{+-}(\bar{3}3)_{c}\left(\xi_{1},\xi_{2}\right)}
\end{array}$ & $\left(\begin{array}{cc}
2848 & -27\\
-27 & 2942
\end{array}\right)$ & $\left(\begin{array}{c}
2841\\
2949
\end{array}\right)$ & $\left(\begin{array}{c}
\left(\begin{array}{cc}
-0.97 & -0.26\end{array}\right)\\
\left(\begin{array}{cc}
-0.26 & 0.97\end{array}\right)
\end{array}\right)$\tabularnewline
$0^{++}$ & $\begin{array}{l}
2^{1}S_{0^{++}(6\bar{6})_{c}\left(\xi_{1},\xi_{2}\right)}\\
2^{1}S_{0^{++}(\bar{3}3)_{c}\left(\xi_{1},\xi_{2}\right)}\\
2^{1}S_{0^{++}(6\bar{6})_{c}\left(\xi_{3}\right)}\\
2^{1}S_{0^{++}(\bar{3}3)_{c}\left(\xi_{3}\right)}
\end{array}$ & $\left(\begin{array}{cccc}
2859 & -53 & -61 & -18\\
-53 & 2903 & -20 & -49\\
-61 & -20 & 3218 & -40\\
-18 & -49 & -40 & 2856
\end{array}\right)$ & $\left(\begin{array}{c}
2781\\
2876\\
2948\\
3232
\end{array}\right)$ & $\left(\begin{array}{c}
\left(\begin{array}{cccc}
-0.61 & -0.52 & -0.16 & -0.57\end{array}\right)\\
\left(\begin{array}{cccc}
0.67 & 0.02 & 0.03 & -0.74\end{array}\right)\\
\left(\begin{array}{cccc}
0.39 & -0.85 & 0.07 & 0.34\end{array}\right)\\
\left(\begin{array}{cccc}
0.15 & 0.02 & -0.98 & 0.09\end{array}\right)
\end{array}\right)$\tabularnewline
$1^{+-}$ & $\begin{array}{l}
2^{3}S_{1^{+-}(\bar{3}3)_{c}\left(\xi_{1},\xi_{2}\right)}\\
2^{3}S_{1^{+-}(\bar{3}3)_{c}\left(\xi_{3}\right)}
\end{array}$ & $\left(\begin{array}{cc}
2920 & -44\\
-44 & 2867
\end{array}\right)$ & $\left(\begin{array}{c}
2842\\
2945
\end{array}\right)$ & $\left(\begin{array}{c}
\left(\begin{array}{cc}
-0.49 & -0.87\end{array}\right)\\
\left(\begin{array}{cc}
-0.87 & 0.49\end{array}\right)
\end{array}\right)$\tabularnewline
$1^{++}$ & $\begin{array}{l}
2^{3}S_{1^{++}(\bar{3}3)_{c}\left(\xi_{1},\xi_{2}\right)}\end{array}$ & $\left(2954\right)$ & 2954 & 1\tabularnewline
$2^{+-}$ & $\begin{array}{l}
2^{5}S_{2^{+-}(\bar{3}3)_{c}\left(\xi_{1},\xi_{2}\right)}\end{array}$ & $\left(2977\right)$ & 2977 & 1\tabularnewline
$2^{++}$ & $\begin{array}{l}
2^{5}S_{2^{++}(\bar{3}3)_{c}\left(\xi_{1},\xi_{2}\right)}\\
2^{5}S_{2^{++}(\bar{3}3)_{c}\left(\xi_{3}\right)}
\end{array}$ & $\left(\begin{array}{cc}
2952 & -28\\
-28 & 2888
\end{array}\right)$ & $\left(\begin{array}{c}
2878\\
2963
\end{array}\right)$ & $\left(\begin{array}{c}
\left(\begin{array}{cc}
-0.35 & -0.94\end{array}\right)\\
\left(\begin{array}{cc}
-0.94 & 0.35\end{array}\right)
\end{array}\right)$\tabularnewline
\hline
\end{tabular}
\end{center}
\end{table*}

\begin{table*}[htp]
\begin{center}
\caption{\label{statesa} The average contributions of each part of the Hamiltonian to the $ss\bar{s}\bar{s}$ configurations with the HOEM. $\langle T\rangle$ stands for the contribution of the kinetic energy term. $\langle V^{Lin}\rangle$ and $\langle V^{Coul}\rangle$ stand for the contributions from the linear confinement potential and Coulomb type potential, respectively. $\langle V^{SS}\rangle$, $\langle V^{T}\rangle$, and $\langle V^{LS}\rangle$ stand for the contributions from the spin-spin interaction term, the tensor potential term, and the spin-orbit interaction term, respectively. The second number in every column is calculated with the GEM. }
\begin{tabular}{clccccccc}
\hline
$J^{P(C)}$ & Configuration & Mass & $\langle T\rangle$ & $\langle V^{Lin}\rangle$ & $\langle V^{Coul}\rangle$ & $\langle V^{SS}\rangle$ & $\langle V^{T}\rangle$ & $\langle V^{LS}\rangle$\tabularnewline
\hline
\multirow{2}{*}{$0^{++}$} & $1^{1}S_{0^{++}(6\bar{6})_{c}}$ & 2365 & 807 & 930 & -774 & 40.75 &  & \tabularnewline
 & $1^{1}S_{0^{++}(\bar{3}3)_{c}}$ & 2293 & 884 & 890 & -812 & -30.29 &  & \tabularnewline
 &  &  &  &  &  &  &  & \tabularnewline
$1^{+-}$ & $\begin{array}{c}
1^{3}S_{1^{+-}(\bar{3}3)_{c}}\end{array}$ & 2323 & 851 & 906 & -797 & 0 &  & \tabularnewline
 &  &  &  &  &  &  &  & \tabularnewline
$2^{++}$ & $\begin{array}{c}
1^{5}S_{2^{++}(\bar{3}3)_{c}}\end{array}$ & 2378 & 793 & 937 & -767 & 53.2 &  & \tabularnewline
\hline
\multirow{2}{*}{$0^{--}$} & $1^{3}P_{0^{--}(6\bar{6})_{c}\left(\xi_{1},\xi_{2}\right)}$ & 2635 & 827 & 1077 & -660 & 4.42 & 34.94 & -11.65\tabularnewline
 & $1^{3}P_{0^{--}(\bar{3}3)_{c}\left(\xi_{1},\xi_{2}\right)}$ & 2694 & 902 & 1093 & -644 & -0.95 & 9.02 & -27.06\tabularnewline
 &  &  &  &  &  &  &  & \tabularnewline
\multirow{3}{*}{$0^{-+}$} & $1^{3}P_{0^{-+}(6\bar{6})_{c}\left(\xi_{1},\xi_{2}\right)}$ & 2616 & 863 & 1056 & -675 & 26.94 & -4.2 & -12.6\tabularnewline
 & $1^{3}P_{0^{-+}(\bar{3}3)_{c}\left(\xi_{1},\xi_{2}\right)}$ & 2685 & 922 & 1083 & -652 & 8.3 & -9.39 & -28.16\tabularnewline
 & $1^{3}P_{0^{-+}(\bar{3}3)_{c}\left(\xi_{3}\right)}$ & 2576 & 976 & 1015 & -703 & 9.96 & -20.88 & -62.65\tabularnewline
 &  &  &  &  &  &  &  & \tabularnewline
\multirow{5}{*}{$1^{--}$} & $1^{3}P_{1^{--}(6\bar{6})_{c}\left(\xi_{1},\xi_{2}\right)}$ & 2585 & 895 & 1037 & -688 & 5.06 & -20.12 & -6.71\tabularnewline
 & $1^{3}P_{1^{--}(\bar{3}3)_{c}\left(\xi_{1},\xi_{2}\right)}$ & 2694 & 902 & 1093 & -644 & -0.95 & -4.51 & -13.53\tabularnewline
 & $1^{5}P_{1^{--}(\bar{3}3)_{c}\left(\xi_{3}\right)}$ & 2584 & 972 & 1018 & -702 & 43.08 & -14.6 & -93.84\tabularnewline
 & $1^{1}P_{1^{--}(\bar{3}3)_{c}\left(\xi_{3}\right)}$ & 2636 & 877 & 1068 & -665 & -5.24 & 0 & 0\tabularnewline
 & $1^{1}P_{1^{--}(6\bar{6})_{c}\left(\xi_{3}\right)}$ & 2889 & 848 & 1210 & -564 & 33.29 & 0 & 0\tabularnewline
 &  &  &  &  &  &  &  & \tabularnewline
\multirow{3}{*}{$1^{-+}$} & $1^{3}P_{1^{-+}(6\bar{6})_{c}\left(\xi_{1},\xi_{2}\right)}$ & 2628 & 845 & 1066 & -668 & 26.34 & 2.03 & -6.08\tabularnewline
 & $1^{3}P_{1^{-+}(\bar{3}3)_{c}\left(\xi_{1},\xi_{2}\right)}$ & 2712 & 882 & 1106 & -637 & 8.27 & 4.33 & -13\tabularnewline
 & $1^{3}P_{1^{-+}(\bar{3}3)_{c}\left(\xi_{3}\right)}$ & 2633 & 884 & 1064 & -668 & 9.08 & 8.73 & -26.19\tabularnewline
 &  &  &  &  &  &  &  & \tabularnewline
\multirow{3}{*}{$2^{--}$} & $1^{3}P_{2^{--}(6\bar{6})_{c}\left(\xi_{1},\xi_{2}\right)}$ & 2620 & 845 & 1066 & -667 & 4.59 & 3.63 & 6.05\tabularnewline
 & $1^{3}P_{2^{--}(\bar{3}3)_{c}\left(\xi_{1},\xi_{2}\right)}$ & 2725 & 859 & 1119 & -628 & -0.58 & 0.83 & 12.4\tabularnewline
 & $1^{5}P_{2^{--}(\bar{3}3)_{c}\left(\xi_{3}\right)}$ & 2665 & 846 & 1087 & -653 & 35.77 & 11.33 & -24.27\tabularnewline
 &  &  &  &  &  &  &  & \tabularnewline
\multirow{3}{*}{$2^{-+}$} & $1^{3}P_{2^{-+}(6\bar{6})_{c}\left(\xi_{1},\xi_{2}\right)}$ & 2638 & 833 & 1074 & -662 & 25.88 & -0.39 & 5.92\tabularnewline
 & $1^{3}P_{2^{-+}(\bar{3}3)_{c}\left(\xi_{1},\xi_{2}\right)}$ & 2733 & 854 & 1123 & -626 & 8.23 & -0.82 & 12.29\tabularnewline
 & $1^{3}P_{2^{-+}(\bar{3}3)_{c}\left(\xi_{3}\right)}$ & 2673 & 830 & 1096 & -646 & 8.53 & -1.56 & 23.44\tabularnewline
 &  &  &  &  &  &  &  & \tabularnewline
$3^{--}$ & $1^{5}P_{3^{--}(\bar{3}3)_{c}\left(\xi_{3}\right)}$ & 2719 & 780 & 1131 & -625 & 31.94 & -2.80 & 41.99\tabularnewline
\hline
\multirow{2}{*}{$0^{+-}$} & $2^{1}S_{0^{+-}(6\bar{6})_{c}\left(\xi_{1},\xi_{2}\right)}$ & 2848/ 2927 & 716/ 844 & 1122/ 1225 & -366/ -529 & 14.48/ 24.74\tabularnewline
 & $2^{1}S_{0^{+-}(\bar{3}3)_{c}\left(\xi_{1},\xi_{2}\right)}$ & 2942/ 2931 & 934/ 941 & 1247/ 1245 & -598/ -618 & -4.08/ 0.8\tabularnewline
 &  &  &  &  &  &  &  & \tabularnewline
\multirow{4}{*}{$0^{++}$} & $2^{1}S_{0^{++}(6\bar{6})_{c}\left(\xi_{1},\xi_{2}\right)}$ & 2858/ 2874 & 839/ 863 & 1181/ 1206 & -548/ -585 & 24.54/ 27.9\tabularnewline
 & $2^{1}S_{0^{++}(\bar{3}3)_{c}\left(\xi_{1},\xi_{2}\right)}$ & 2903/ 2918 & 956/ 954 & 1236/ 1235 & -639/ -620 & -12.68/ -12.4\tabularnewline
 & $2^{1}S_{0^{++}(6\bar{6})_{c}\left(\xi_{3}\right)}$ & 3216/ 3148 & 898/ 922 & 1394/ 1364 & -465/ -517 & 26.81/ 17.52\tabularnewline
 & $2^{1}S_{0^{++}(\bar{3}3)_{c}\left(\xi_{3}\right)}$ & 2855/ 2841 & 891/ 899 & 1189/ 1197 & -583/ -622 & -3.65/ 4.74\tabularnewline
 &  &  &  &  &  &  &  & \tabularnewline
\multirow{2}{*}{$1^{+-}$} & $2^{3}S_{1^{+-}(\bar{3}3)_{c}\left(\xi_{1},\xi_{2}\right)}$ & 2919/ 2934 & 937/ 926 & 1247/ 1252 & -631/ -610 & 4.36/ 3.51\tabularnewline
 & $2^{3}S_{1^{+-}(\bar{3}3)_{c}\left(\xi_{3}\right)}$ & 2866/ 2851 & 877/ 895 & 1195/ 1201 & -575/ -621 & 7.29/ 14.61\tabularnewline
 &  &  &  &  &  &  &  & \tabularnewline
$1^{++}$ & $2^{3}S_{1^{++}(\bar{3}3)_{c}\left(\xi_{1},\xi_{2}\right)}$ & 2954/ 2943 & 919/ 926 & 1256/ 1256 & -591/ -614 & 7.95/ 12.47\tabularnewline
 &  &  &  &  &  &  &  & \tabularnewline
$2^{+-}$ & $2^{5}S_{2^{+-}(\bar{3}3)_{c}\left(\xi_{1},\xi_{2}\right)}$ & 2976/ 2965 & 890/ 905 & 1272/ 1271 & -577/ -607 & 30.03/ 34.34\tabularnewline
 &  &  &  &  &  &  &  & \tabularnewline
\multirow{2}{*}{$2^{++}$} & $2^{5}S_{2^{++}(\bar{3}3)_{c}\left(\xi_{1},\xi_{2}\right)}$ & 2952/ 2964 & 902/ 898 & 1268/ 1272 & -615/ -601 & 35.6/ 33.36\tabularnewline
 & $2^{5}S_{2^{++}(\bar{3}3)_{c}\left(\xi_{3}\right)}$ & 2887/ 2871 & 851/ 868 & 1207/ 1220 & -560/ -612 & 27.25/ 33.3\tabularnewline
\hline
\end{tabular}
\end{center}
\end{table*}

\begin{figure*}
\centering \epsfxsize=16.8 cm \epsfbox{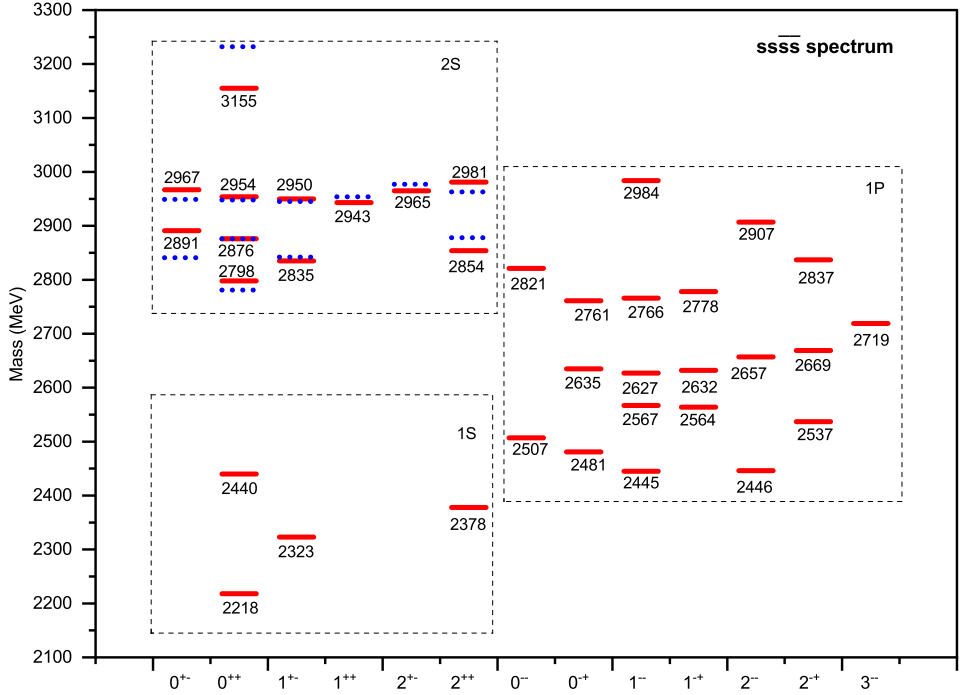} \vspace{-1.4 cm} \caption{Mass spectrum for the $ss\bar{s}\bar{s}$ system.
The solid and dotted lines stand for the results predicted by the GEM and HOEM, respectively.}\label{figs}
\end{figure*}

\begin{table*}[htp]
\begin{center}
\caption{\label{GEMss} Predicted mass spectrum for the $2S$-wave $ss\bar{s}\bar{s}$ system with the GEM.}
\scalebox{1.0}{
\begin{tabular}{ccccc}
\hline
$J^{P(C)}$ & Configuration & $\langle H\rangle$ (MeV) & Mass (MeV) & Eigenvector\tabularnewline
\hline
$0^{+-}$ & $\begin{array}{l}
2^{1}S_{0^{+-}(6\bar{6})_{c}\left(\xi_{1},\xi_{2}\right)}\\
2^{1}S_{0^{+-}(\bar{3}3)_{c}\left(\xi_{1},\xi_{2}\right)}
\end{array}$ & $\left(\begin{array}{cc}
2927 & 38\\
38 & 2931
\end{array}\right)$ & $\left(\begin{array}{c}
2891\\
2967
\end{array}\right)$ & $\left(\begin{array}{cc}
-0.73 & 0.69\\
0.69 & 0.73
\end{array}\right)$\tabularnewline
$0^{++}$ & $\begin{array}{l}
2^{1}S_{0^{++}(6\bar{6})_{c}\left(\xi_{1},\xi_{2}\right)}\\
2^{1}S_{0^{++}(\bar{3}3)_{c}\left(\xi_{1},\xi_{2}\right)}\\
2^{1}S_{0^{++}(6\bar{6})_{c}\left(\xi_{3}\right)}\\
2^{1}S_{0^{++}(\bar{3}3)_{c}\left(\xi_{3}\right)}
\end{array}$ & $\left(\begin{array}{cccc}
2874 & -48 & 27 & -24\\
-48 & 2918 & -8 & -42\\
27 & -8 & 3148 & -31\\
-24 & -42 & -31 & 2841
\end{array}\right)$ & $\left(\begin{array}{c}
2798\\
2876\\
2954\\
3155
\end{array}\right)$ & $\left(\begin{array}{cccc}
-0.50 & -0.46 & -0.04 & -0.73\\
0.74 & 0.20 & -0.14 & -0.62\\
0.43 & -0.87 & -0.06 & 0.25\\
-0.11 & 0.04 & -0.99 & 0.10
\end{array}\right)$\tabularnewline
$1^{+-}$ & $\begin{array}{l}
2^{3}S_{1^{+-}(\bar{3}3)_{c}\left(\xi_{1},\xi_{2}\right)}\\
2^{3}S_{1^{+-}(\bar{3}3)_{c}\left(\xi_{3}\right)}
\end{array}$ & $\left(\begin{array}{cc}
2934 & -40\\
-40 & 2851
\end{array}\right)$ & $\left(\begin{array}{c}
2835\\
2950
\end{array}\right)$ & $\left(\begin{array}{cc}
-0.38 & -0.93\\
-0.93 & 0.38
\end{array}\right)$\tabularnewline
$1^{++}$ & $\begin{array}{l}
2^{3}S_{1^{++}(\bar{3}3)_{c}\left(\xi_{1},\xi_{2}\right)}\end{array}$ & $\left(2943\right)$ & 2943 & 1\tabularnewline
$2^{+-}$ & $\begin{array}{l}
2^{5}S_{2^{+-}(\bar{3}3)_{c}\left(\xi_{1},\xi_{2}\right)}\end{array}$ & $\left(2965\right)$ & 2965 & 1\tabularnewline
$2^{++}$ & $\begin{array}{l}
2^{5}S_{2^{++}(\bar{3}3)_{c}\left(\xi_{1},\xi_{2}\right)}\\
2^{5}S_{2^{++}(\bar{3}3)_{c}\left(\xi_{3}\right)}
\end{array}$ & $\left(\begin{array}{cc}
2964 & -43\\
-43 & 2871
\end{array}\right)$ & $\left(\begin{array}{c}
2854\\
2981
\end{array}\right)$ & $\left(\begin{array}{cc}
-0.36 & -0.93\\
-0.93 & 0.36
\end{array}\right)$\tabularnewline
\hline
\end{tabular}}
\end{center}
\end{table*}

\section{Results and discussions}\label{Results}

Our predictions of the $T_{ss\bar{s}\bar{s}}$ mass spectrum with the HOEM are given in Table \ref{mass of ssss}, where the
components of different configurations for a physical state can be seen.
For example, the two $0^{++}$ ground states are mixing states between two different configurations
$^{1}S_{0^{++}(6\bar{6})_{c}}$ and $^{1}S_{0^{++}(\bar{3}3)_{c}}$ due to a strong contribution
of the confinement potential to the non-diagonal elements. To see the contributions from
each part of the Hamiltonian to the mass of different configurations, we also present our results in Table~\ref{statesa}.
It is found that both the kinetic energy term $\langle T\rangle$ and the linear confinement potential term $\langle V^{Lin}\rangle$
contribute a large positive value to the mass, while the Coulomb type potential $\langle V^{Coul}\rangle$
has a large cancelation with these two terms. The spin-spin interaction term $\langle V^{SS}\rangle$,
the tensor potential term $\langle V^{T}\rangle$, and/or the spin-orbit interaction term $\langle V^{LS}\rangle$
have also sizeable contributions to some configurations. Thus, as a reliable calculation, both the spin-independent and
spin-dependent potentials should be reasonably included for the $ss\bar{s}\bar{s}$ system.
For clarity, our predicted $T_{ss\bar{s}\bar{s}}$ spectrum is plotted in Fig.~\ref{figs}.

\subsection{Discussions of the numerical method}\label{aaa}

Herein we discuss the differences of numerical results between
the expansion method with the harmonic oscillator wave functions (HOEM) used in present work
and the Gaussian expansion method (GEM) often adopted in the literature.
For the $1S$-, $1P$-wave $T_{(ss\bar{s}\bar{s})}$ states, etc., there are no radial excitations. Thus,
the GEM is the same as the HOEM. For the first radial excited $2S$-wave $T_{(ss\bar{s}\bar{s})}$ states,
the HOEM is different from the GEM because the trail harmonic oscillator wave functions
are different from the Gaussian functions.

To see the differences between the two expansion methods we also give our predictions of
the $2S$-wave $T_{(ss\bar{s}\bar{s})}$ states based on the GEM.
It should be mentioned that by fully expanding $\prod_{i=1}^3 R_{n_{\xi_i} l_{\xi_i}}({\xi_i})$
with the GEM, one cannot distinguish the $\xi_1$ and $\xi_2$ excited modes which are
defined for the $2S$ configurations presented in Table~\ref{states}. Then we cannot
numerically work out the masses for the following states of $0^{+-}$ ($2^{1}S_{0^{+-}(6\bar{6})_{c}\left(\xi_{1},\xi_{2}\right)}$ and $2^{1}S_{0^{+-}(\bar{3}3)_{c}\left(\xi_{1},\xi_{2}\right)}$), $1^{++}$ ($2^{3}S_{1^{++}(\bar{3}3)_{c}\left(\xi_{1},\xi_{2}\right)}$), and $2^{+-}$ ($2^{5}S_{2^{+-}(\bar{3}3)_{c}\left(\xi_{1},\xi_{2}\right)}$) listed in Table~\ref{states}.
To overcome this problem, the spatial wave functions containing the radial excitations are expanded with the Gaussian functions,
while the spatial wave functions containing no excitations are adopted the single
Gaussian function as an approximation. We have tested the single Gaussian approximation in the calculations
of the ground $1S$ $T_{(ss\bar{s}\bar{s})}$ states, the numerical values are
reasonably consistent with those calculated with a series of Gaussian functions.
The differences of the numerical results between these two methods are about 10 MeV.

Our numerical results for the $2S$-wave $T_{(ss\bar{s}\bar{s})}$ states with
the GEM are listed in Table~\ref{statesa} and Table~\ref{GEMss}.
From Table~\ref{statesa}, it is found that the numerical values for the $0^{+-}$
configuration $2^{1}S_{0^{+-}(6\bar{6})_{c}\left(\xi_{1},\xi_{2}\right)}$ and $0^{++}$
configuration $2^{1}S_{0^{++}(6\bar{6})_{c}\left(\xi_{3}\right)}$ calculated with
the HOEM are significantly different from those obtained with the GEM.
For these two configurations, the predicted mass differences by the HOEM and GEM can reach up to $\sim 70$ MeV. However, for the other
$2S$-wave $T_{(ss\bar{s}\bar{s})}$ configurations the numerical values of these two methods
are comparable with each other. The differences of the predicted masses
between these two methods are about $10-20$ MeV. It should be mentioned that
the Coulomb type potential $\langle V^{Coul}\rangle$ for the
$2S$-wave states seems to be sensitive to the numerical methods as shown in Table~\ref{statesa}.

In brief, most of the predictions are consistent with each other between the HOEM and GEM. The uncertainties from the numerical methods
do not change our main predictions of the $T_{(ss\bar{s}\bar{s})}$ spectrum,
Although some numerical results for the $J^{PC}=0^{+-}$ and $0^{++}$ $2S$ states
show a significant numerical method dependence (See Fig.~\ref{figs}), the GEM may give a slightly more accurate
numerical result based on our tests of the charmonium spectrum. In the following, our discussions
of the $2S$ states are based on the GEM calculations.

\subsection{$S$-wave states}

There are four $1S$-wave $T_{ss\bar{s}\bar{s}}$ states with $J^{PC}=0^{++},\ 1^{+-}, \ 2^{++}$ in the quark model. Their masses
are predicted to be in the range of $2.21-2.44$ GeV. In contrast, the $2S$ wave includes twelve states. Except for the highest mass state $T_{(ss\bar{s}\bar{s})0^{++}}(3155)$, their masses lie in a relative narrow range of  $2.78-2.98$ GeV.
Apart from the conventional quantum numbers, i.e., $J^{PC}=0^{++}, \ 1^{++}, \ 1^{+-}, \ 2^{++}$, the $2S$-wave
can access exotic quantum numbers, i.e., $J^{PC}=0^{+-}, \ 2^{+-}$.

\subsubsection{$0^{++}$ states}

In the $1S$-wave mutiplets, the two $0^{++}$ ground states include $T_{(ss\bar{s}\bar{s})0^{++}}(2218)$ and
$T_{(ss\bar{s}\bar{s})0^{++}}(2440)$. Their mass splitting reaches up to about $200$ MeV.
These two states have a strong mixing between the two color structures $|6\bar{6}\rangle_c$ and $|\bar{3}3\rangle_c$.
Their masses are much larger than the mass threshold of $\phi\phi$. Thus, they may easily decay into
$\phi\phi$ pair through quark rearrangements. The mass of the lowest $0^{++}$ $T_{ss\bar{s}\bar{s}}$ in our model
is close to the prediction of $2203$ MeV in the relativistic diquark-antidiquark model~\cite{Ebert:2008id}.
However, it turns out to be much higher than the predicted value $1716$ MeV by the relativized quark model
with a diquark-antidiquark approximation~\cite{Lu:2019ira}. There might be some crucial dynamics missing in the diquark-antidiquark approximation. As a test of the  diquark-antidiquark approximation we adopt the
approximation as done in Ref.~\cite{Lu:2019ira} and  calculate the mass of the $0^{++}$ $T_{ss\bar{s}\bar{s}}$ state
with the same potential model parameters. We obtain a mass of $1758$ MeV,
which is comparable with the prediction of Ref.~\cite{Lu:2019ira}, but
is obviously smaller than the results without the diquark-antidiquark approximation.

In the  $2S$-wave sector, there are four $0^{++}$ states, $T_{(ss\bar{s}\bar{s})0^{++}}(2798)$,
$T_{(ss\bar{s}\bar{s})0^{++}}(2876)$, $T_{(ss\bar{s}\bar{s})0^{++}}(2954)$, and $T_{(ss\bar{s}\bar{s})0^{++}}(3155)$,
predicted in the NRPQM with GEM. A strong mixing
between the two color structures $|6\bar{6}\rangle_c$ and $|\bar{3}3\rangle_c$ is also found among these states. In particular,
the radial excitation modes $(\xi_1,\xi_2)$ and $(\xi_3)$ strongly mix with each other.
The highest state $T_{(ss\bar{s}\bar{s})0^{++}}(3155)$ is nearly a pure configuration of
$^{1}S_{0^{++}(6\bar{6})_{c}\left(\xi_{3}\right)}$, with the color structure
$|6\bar{6}\rangle_c$ and the radial excitation
between diquark $(ss)$ and anti-diquark $(\bar{s}\bar{s})$. The special color structure
of $T_{(ss\bar{s}\bar{s})0^{++}}(3155)$ leads to a rather large mass gap $\Delta\simeq 201$ MeV from the nearby  $T_{(ss\bar{s}\bar{s})0^{++}}(2954)$.
These $2S$-wave $0^{++}$ $ss\bar{s}\bar{s}$ states may easily decay into
$\phi\phi$, $\phi\phi(1680)$ final states through quark rearrangements.
They may also easily decay into $K_s^0K_s^0$ and $K^+K^-$ final states through the $s\bar{s}$ annihilation and a pair of nonstrange $q\bar{q}$ creation. One also notices that these states may directly decay into $\Xi\bar{\Xi}$ baryon pair with a light $q\bar{q}$ pair creation.

Some evidences for $T_{(ss\bar{s}\bar{s})0^{++}}(2218)$ and
$T_{(ss\bar{s}\bar{s})0^{++}}(2440)$ may have been seen in the previous experiments. Recently,
Kozhevnikov carried out a dynamical analysis of the resonance contributions to $J/\psi\to \gamma X\to \gamma \phi\phi$~\cite{Kozhevnikov:2019lmy} with the data from BESIII~\cite{Ablikim:2016hlu}. Two $0^{++}$ resonances with masses at $\sim 2.2$ GeV and
$\sim2.4$ GeV, were extracted from the data. Evidence for a scalar around $2.2$ GeV in the $\phi\phi$ mass spectra in $B_s^0\to J/\psi \phi\phi$~\cite{Aaij:2016qim} was also reported by Ref.~\cite{Kozhevnikov:2017nlr}.
Considering the mass and decay mode, these two scalar structures may be good candidates for $T_{(ss\bar{s}\bar{s})0^{++}}(2218)$ and $T_{(ss\bar{s}\bar{s})0^{++}}(2440)$.

It should be mentioned that $f_0(2200)$ is listed in RPP~\cite{Tanabashi:2018oca} as a well-established state. It has been seen in the $K_s^0K_s^0$, $K^+K^-$ and $\eta\eta$, and may be assigned to $T_{(ss\bar{s}\bar{s})0^{++}}(2218)$. Some qualitative features can be expected: (i) The $0^{++}$ $T_{ss\bar{s}\bar{s}}$ state can decay into $\eta\eta$, $\eta'\eta'$, and $\eta\eta'$ through quark rearrangements via the $s\bar{s}$ component in the $\eta$ and $\eta'$ mesons. An approximate branching ratio fraction can be examined: $BR(\eta\eta) : BR(\eta'\eta'): BR(\eta\eta')\simeq \sin^4\alpha_P : \cos^4\alpha_P : 2\sin^2\alpha_P\cos^2\alpha_P\simeq 0.24 : 0.25 : 0.50$, with $\alpha_P\equiv\arctan\sqrt{2}+\theta_P\simeq 44.7^\circ$ and without including the phase space factors. (ii) The $0^{++}$ states may also easily decay into $K_s^0K_s^0$ and $K^+K^-$ final states through annihilating a pair of $s\bar{s}$ and creating a pair of light $q\bar{q}$. (iii) It is interesting to note that no conventional $0^{++}$ $s\bar{s}$ states are predicted around 2.2 GeV in most literatures~\cite{Li:2020xzs}.

To establish the $0^{++}$ ground states $T_{(ss\bar{s}\bar{s})0^{++}}(2218)$ and $T_{(ss\bar{s}\bar{s})0^{++}}(2440)$, a combined study of decay channels, such as $\phi\phi$, $K_s^0K_s^0$, $K^+K^-$, $\eta\eta$, $\eta'\eta'$, and $\eta\eta'$, should be necessary. The $2S$-wave $0^{++}$ states can be probed in these meson pair decay channels including higher channels such as $\phi\phi(1680)$, and some baryon pair decay channels such as  $\Xi\bar{\Xi}$.

\subsubsection{$2^{++}$ states}

There is only one $2^{++}$ state $T_{(ss\bar{s}\bar{s})2^{++}}(2378)$ in the $1S$-wave states.
This state lies between the two $0^{++}$ ground states, and has a pure
$|\bar{3}3\rangle_c$ color structure. $T_{(ss\bar{s}\bar{s})2^{++}}(2378)$ may have
large decay rates into the $\phi\phi$, $\eta\eta$ and $\eta'\eta'$ final states through
quark rearrangements, and/or into $K^{(*)}\bar{K}^{(*)}$ final states through the annihilation of $s\bar{s}$ and creation of a pair of nonstrange $q\bar{q}$. It should be mentioned that with the diquark-antidiquark approximation,
the mass of the $2^{++}$ state is predicted to be 2192 MeV, which is about 200 MeV lower than the four-body calculation results.

The $f_2(2340)$ resonance listed in RPP~\cite{Tanabashi:2018oca} may be assigned to $T_{(ss\bar{s}\bar{s})2^{++}}(2378)$.
Besides the measured mass $2345^{+50}_{-40}$ MeV, the observed decay modes $\phi\phi$ and $\eta\eta$
are consistent with the expectation of the tetraquark scenario. On the other hand, as a conventional $s\bar{s}$ state the $f_2(2340)$ cannot be easily accommodated by the quark model expectation~\cite{Li:2020xzs}.
The relativistic quark model calculation of Ref.~\cite{Ebert:2008id}
also supports the $f_2(2340)$ to be assigned as the $T_{ss\bar{s}\bar{s}}$ ground state with $2^{++}$.
To confirm this assignment, the other main decay modes of $T_{(ss\bar{s}\bar{s})2^{++}}(2378)$ such as $\eta\eta'$, $\eta'\eta'$, $K^{(*)}\bar{K}^{(*)}$ should be investigated in experiment.

For the  $2S$-wave sector, there are two $2^{++}$ $ss\bar{s}\bar{s}$ states $T_{(ss\bar{s}\bar{s})2^{++}}(2854)$ and
$T_{(ss\bar{s}\bar{s})2^{++}}(2981)$ predicted in our model, which are dominated by the
$^{5}S_{2^{++}(\bar{3}3)_{c}\left(\xi_{3}\right)}$ and $^{5}S_{2^{++}(\bar{3}3)_{c}\left(\xi_{1},\xi_{2}\right)}$
configurations, respectively. Their masses are predicted to be above the thresholds of
$\phi\phi$, $\phi\phi(1680)$ and $\Xi(1530)\bar{\Xi}$. Therefore, experimental search for their signals in these decay channels should be helpful for understanding these tensor tetraquarks.

%Observations of the reactions $e^+e^-\to \gamma X$, $X\to \phi\phi,\Xi(1530)\bar{\Xi}$ at BESIII may find some evidences of these high $2^{++}$ $T_{ss\bar{s}\bar{s}}$ states.

\subsubsection{$1^{+-}$ states}

In the $1S$-wave multiplets $T_{(ss\bar{s}\bar{s})1^{+-}}(2323)$ is the only state with $C=-1$, and has a pure
$|\bar{3}3\rangle_c$ color structure. Its mass is about 100 MeV larger than the lowest $1S$-wave state $T_{(ss\bar{s}\bar{s})0^{++}}(2218)$.
Its mass is about $200\sim 300$ MeV larger than that predicted by the QCD sum rules
~\cite{Cui:2019roq} and the relativized quark model~\cite{Lu:2019ira} in the diquark picture.
The mass of the $1^{+-}$ state may be notably underestimated in the diquark picture.
As a comparison we calculate the mass of the $1^{+-}$ state in the diquark picture using the
same potential model parameter set adopted in present work, and obtain a mass of $1936$ MeV, which
is 300 MeV smaller than the NRQPM prediction 2236 MeV.

$T_{(ss\bar{s}\bar{s})1^{+-}}(2323)$ may easily decay into $\eta \phi$ and $\eta' \phi$ through the quark rearrangements.
The decay of $\psi'/J/\psi\to \phi\eta\eta'$ can access this state in $\eta\phi$ and $\eta'\phi$ channels.
It should be mentioned that some hints of $T_{(ss\bar{s}\bar{s})1^{+-}}(2323)$ may have been found
in the $\eta' \phi$ invariant mass spectrum around $2.3-2.4$ GeV by observing the $J/\psi\to \phi\eta\eta'$ reaction at BESIII recently~\cite{Ablikim:2018xuz}.

For the  $2S$-wave sector, there are two states, i.e. $1^{+-}$ $ss\bar{s}\bar{s}$ $T_{(ss\bar{s}\bar{s})1^{+-}}(2835)$ and
$T_{(ss\bar{s}\bar{s})1^{+-}}(2950)$ predicted in the quark model. There are sizeable configuration mixings in these two states.
The $T_{(ss\bar{s}\bar{s})1^{+-}}(2835)$ is dominated by the $2^{3}S_{1^{+-}(\bar{3}3)_{c}\left(\xi_{3}\right)}$
configuration, which has a $|\bar{3}3\rangle_c$ color structure, and the radial excitation occurs
between diquark $(ss)$ and anti-diquark $(\bar{s}\bar{s})$ (i.e., the $\xi_3$ mode).
The $T_{(ss\bar{s}\bar{s})1^{+-}}(2950)$ is dominated by the $2^{3}S_{1^{+-}(\bar{3}3)_{c}\left(\xi_{1},\xi_{2}\right)}$
configuration, whose radial excitation occurs in the diquark $(ss)$ and anti-diquark $(\bar{s}\bar{s})$. Apart from the $\eta\phi$ and $\eta'\phi$ decay channels it may favor decays into a pseudoscalar plus a radially excited vector (i.e., $\eta \phi(1680)$ and $\eta' \phi(1680)$), or a radially excited pseudoscalar plus a vector (i.e., $\eta(1295)\phi$ and $\eta(1405)\phi$), through the quark rearrangements.

It should be mentioned that in Refs.~\cite{Cui:2019roq,tetra7} the authors suggest that the new structure $X(2063)$ observed in the
$J/\psi\to \phi\eta\eta'$ at BESIII~\cite{Ablikim:2018xuz} could be a $1^{+-}$ $T_{ss\bar{s}\bar{s}}$ candidate according to
the QCD sum rule calculation. However, the observed mass of $X(2063)$ is too small to be comparable with our
quark model predictions.

\subsubsection{$0^{+-}$ and $2^{+-}$ states}

In the $2S$-wave multiplets, there are two $0^{+-}$ states, $T_{(ss\bar{s}\bar{s})0^{+-}}(2891)$
and $T_{(ss\bar{s}\bar{s})0^{+-}}(2967)$, predicted in the NRPQM with GEM. There is a
strong configuration mixing between $^{1}S_{0^{+-}(6\bar{6})_{c}\left(\xi_{1},\xi_{2}\right)}$ and $^{1}S_{0^{+-}(\bar{3}3)_{c}\left(\xi_{1},\xi_{2}\right)}$. There is only one $2^{+-}$ state $T_{(ss\bar{s}\bar{s})2^{+-}}(2965)$ corresponding
to the configuration $^{5}S_{2^{+-}(\bar{3}3)_{c}\left(\xi_{1},\xi_{2}\right)}$.
The $0^{+-}$ and $2^{+-}$ are exotic quantum numbers which cannot be accommodated by the conventional $q\bar{q}$ scenario.
The $P$-wave decays into the $\eta h_1(1P)$ and $\eta' h_1(1P)$ channels could be useful for the search for these states in experiments.

\subsection{$1P$-wave states}

There are twenty $1P$-wave $T_{ss\bar{s}\bar{s}}$ states predicted in the NRPQM.
Apart from the conventional quantum numbers, i.e., $J^{PC}=0^{-+}, \ 1^{--}, \ 2^{-+}, \ 2^{--}, \ 3^{--}$, the $P$-wave can access exotic quantum numbers, i.e., $J^{PC}=0^{--}, \ 1^{-+}$. The masses of the $1P$-wave $T_{ss\bar{s}\bar{s}}$ states scatter in a wide range of about $2.4-3.0$ GeV. The masses of the low-lying $1P$-wave states may highly overlap with the heaviest $1S$-wave state $T_{(ss\bar{s}\bar{s})0^{++}}(2440)$.

\subsubsection{$0^{-+}$ states}

There are three $0^{-+}$ states, $T_{(ss\bar{s}\bar{s})0^{-+}}(2481)$,
$T_{(ss\bar{s}\bar{s})0^{-+}}(2635)$, and $T_{(ss\bar{s}\bar{s})0^{-+}}(2761)$, predicted in the NRPQM.
They are mixed states with two color structures $|6\bar{6}\rangle_c$ and $|\bar{3}3\rangle_c$, and
also mixed states between two orbital excitations $(\xi_1,\xi_2)$ and $\xi_3$ modes.
They can decay into $\phi h_1(1P)$ via an $S$ wave, or $\phi\phi$ via a $P$ wave, through the quark rearrangements.

In 2016, the BESIII Collaboration observed a new resonance $X(2500)$ with a mass of
$2470^{+15}_{-19}$$^{+101}_{-23}$ MeV and a width of $230^{+64}_{-35}$$^{+56}_{-33}$ MeV
in $J/\psi\to \gamma \phi\phi$~\cite{Ablikim:2016hlu}. The preferred spin-parity numbers for $X(2500)$
are $J^{PC}=0^{-+}$~\cite{Ablikim:2016hlu}. The $X(2500)$ resonance may be a candidate for
$T_{(ss\bar{s}\bar{s})0^{-+}}(2481)$ in terms of mass, decay modes and quantum
numbers although $X(2500)$ may favor the $4^1S_0$  $s\bar{s}$ state as suggested in
our previous work~\cite{Li:2020xzs}. In the recent work of Ref.~\cite{Dong:2020okt}, the authors also
suggested $X(2500)$ to be a $0^{-+}$ $T_{ss\bar{s}\bar{s}}$ state according to
the QCD sum rule studies. A measurement of the branching fraction of
$\mathcal{B}[X(2500)\to \phi\phi]$ might provide a test of the nature of $X(2500)$. The decay rate of $T_{(ss\bar{s}\bar{s})0^{-+}}(2481)$
into $\phi\phi$ through the quark rearrangements should be significantly larger
than that via an $s\bar{s}$ pair production for the $4^1S_0$  $s\bar{s}$ state.

\subsubsection{$1^{--}$ states}

There are five $1^{--}$ states, $T_{(ss\bar{s}\bar{s})1^{--}}(2445)$,
$T_{(ss\bar{s}\bar{s})1^{--}}(2567)$, $T_{(ss\bar{s}\bar{s})1^{--}}(2627)$, $T_{(ss\bar{s}\bar{s})1^{--}}(2766)$, and $T_{(ss\bar{s}\bar{s})1^{--}}(2984)$, predicted in the NRPQM.
Their masses scatter in a rather wide range of about $2.4-3.0$ GeV.
From Table~\ref{mass of ssss}, it is found that there are obvious configuration mixings in these tetraquark states except that
$T_{(ss\bar{s}\bar{s})1^{--}}(2627)$ may nearly be a pure $^{1}P_{1^{--}(\bar{3}3)_{c}\left(\xi_{3}\right)}$ state.
The lowest state $T_{(ss\bar{s}\bar{s})1^{--}}(2445)$ is dominated by the
$^{3}P_{1^{--}(6\bar{6})_{c}\left(\xi_{1},\xi_{2}\right)}$ configuration. Its orbital excitation
mainly occurs within the diquark $(ss)$ or anti-diquark $(\bar{s}\bar{s})$.
Meanwhile, the highest state $T_{(ss\bar{s}\bar{s})1^{--}}(2984)$ is dominated by the
$^{1}P_{1^{--}(6\bar{6})_{c}\left(\xi_{3}\right)}$ configuration, and the orbital excitation
occurs between the diquark $(ss)$ and anti-diquark $(\bar{s}\bar{s})$.

The vector meson $\phi(2170)$ in RPP~\cite{Tanabashi:2018oca}
is suggested to be a $1^{--}$ $T_{ss\bar{s}\bar{s}}$ state in the literature~\cite{tetra1,tetra2,tetra3,tetra5,tetra6,tetra7}
since it is hard to be explained as a conventional meson state
according to its measured decay modes~\cite{Ablikim:2020pgw,Ablikim:2020coo}.
Furthermore, the $X(2239)$ resonance, which was observed in $e^+e^-\to K^+K^-$  by the BESIII Collaboration~\cite{Ablikim:2018iyx},
was suggested to be a candidate of the lowest $1^{--}$ $T_{ss\bar{s}\bar{s}}$ state by comparing with the
mass spectrum from the relativized quark model~\cite{Lu:2019ira} and the QCD two-point sum rule method~\cite{Azizi:2019ecm}. However, our calculations indicate that neither $\phi(2170)$
nor $X(2239)$ can be assigned to a $1^{--}$ $T_{ss\bar{s}\bar{s}}$ state since their measured masses are
much lower than our predictions. It should be mentioned that in recent studies the $\phi(2170)$ was
considered as a vector tetraquark state with content $su\bar{s}\bar{u}$ rather than as a state
$ss\bar{s}\bar{s}$~\cite{Agaev:2019coa,Agaev:2020zad}.

The $1^{--}$ $T_{ss\bar{s}\bar{s}}$ states may have large decay rates into the $f_0(980)\phi$ via an $S$ wave, or into $\eta\phi$ and $\eta'\phi$ via a $P$ wave. There are some experimental evidences for structures around 2.4 GeV observed in the $f_0(980)\phi$ invariant mass spectrum
from \emph{BABAR}~\cite{Aubert:2006bu,Aubert:2007ur}, Belle~\cite{Shen:2009zze}, BESII~\cite{Ablikim:2007ab}, and BESIII~\cite{Ablikim:2014pfc}, which could be signals of the $1^{--}$ $ss\bar{s}\bar{s}$
tetraquark states~\cite{Chen:2018kuu}.
For the heavier states $T_{(ss\bar{s}\bar{s})1^{--}}(2766)$ and $T_{(ss\bar{s}\bar{s})1^{--}}(2984)$, they can also decay
into $\Xi\bar{\Xi}$ baryon pair through a $q\bar{q}$ pair production in vacuum. Thus, experimental search for these states in $e^+e^-\to \Xi\bar{\Xi}$ should be very interesting.

\subsubsection{$1^{-+}$ states}

There are three $1^{-+}$ states, $T_{(ss\bar{s}\bar{s})1^{-+}}(2564)$,
$T_{(ss\bar{s}\bar{s})1^{-+}}(2632)$, and $T_{(ss\bar{s}\bar{s})1^{-+}}(2778)$, predicted in the NRPQM.
Note that  $1^{-+}$ are exotic quantum numbers which cannot be accommodated by the conventional $q\bar{q}$ scenario.
Both the lowest mass state $T_{(ss\bar{s}\bar{s})1^{-+}}(2564)$ and highest mass state
$T_{(ss\bar{s}\bar{s})1^{-+}}(2778)$ are mixed states between the two color structures
$|6\bar{6}\rangle_c$ and $|\bar{3}3\rangle_c$, and their orbital excitations are dominated by
the $(\xi_1,\xi_2)$ mode. The middle state $T_{(ss\bar{s}\bar{s})1^{-+}}(2632)$
is dominated by the $^{3}P_{1^{-+}(\bar{3}3)_{c}\left(\xi_{3}\right)}$ configuration of which
the orbital excitation mainly occurs between the diquark $(ss)$ and anti-diquark $(\bar{s}\bar{s})$.
It should be noted that a corresponding state in the relativized
quark model~\cite{Lu:2019ira} has a mass of 2581 MeV, which is about 50 MeV smaller than our prediction.

These $1^{-+}$ $T_{ss\bar{s}\bar{s}}$ states may easily decay into $\phi h_1(1P)$, $\eta^{(')} f_1(1420)$,
$\phi\phi$ channels through the quark rearrangements. They can be searched for in $\chi_{cJ}(1P)\to \eta \phi\phi, \phi K^*K$ with sufficient $\chi_{cJ}(1P)$ data samples at BESIII, although no obvious structures were found in previous
observations~\cite{Ablikim:2019yjw,Ablikim:2015lnn}.

\subsubsection{$2^{-+}$ states}

There are three $2^{-+}$ states, $T_{(ss\bar{s}\bar{s})2^{-+}}(2537)$,
$T_{(ss\bar{s}\bar{s})2^{-+}}(2669)$, and $T_{(ss\bar{s}\bar{s})2^{-+}}(2837)$, predicted in the NRPQM.
Both the lowest mass state $T_{(ss\bar{s}\bar{s})2^{-+}}(2537)$ and highest mass state
$T_{(ss\bar{s}\bar{s})2^{-+}}(2837)$ are mixed states between the two color structures
$|6\bar{6}\rangle_c$ and $|\bar{3}3\rangle_c$. Their orbital excitations are dominated by
the $(\xi_1,\xi_2)$ mode. The middle state $T_{(ss\bar{s}\bar{s})2^{-+}}(2669)$
is dominated by the $^{3}P_{2^{-+}(\bar{3}3)_{c}\left(\xi_{3}\right)}$ configuration of which the orbital excitation occurs between the diquark $(ss)$ and anti-diquark $(\bar{s}\bar{s})$.
A corresponding state in the relativized
quark model~\cite{Lu:2019ira} has a mass of 2619 MeV, which is about 50 MeV smaller than our prediction.

These $2^{-+}$ $T_{ss\bar{s}\bar{s}}$ states may easily fall apart into $\phi h_1(1P)$ and $\eta f_2'(1525)$ in an $S$ wave, or into
$\phi\phi$ in a $P$ wave through the quark rearrangements. For the high mass state $T_{(ss\bar{s}\bar{s})2^{-+}}(2837)$,
the strong decay mode $\Xi\bar{\Xi}$ also opens. These states can be searched for in $\chi_{c2}(1P)\to \eta T_{(ss\bar{s}\bar{s})2^{-+}}\to \eta \eta f_2'(1525)\to \eta \eta K \bar{K}$ at BESIII with the sufficient $\chi_{c2}(1P)$ data samples.

\subsubsection{$0^{--}$ states}

There are two states with exotic quantum numbers of $0^{--}$, $T_{(ss\bar{s}\bar{s})0^{--}}(2507)$ and
$T_{(ss\bar{s}\bar{s})0^{--}}(2821)$, predicted in the NRPQM.
These two states have a strong mixing between the two color structures $|6\bar{6}\rangle_c$ and $|\bar{3}3\rangle_c$.
The orbital excitation is the $(\xi_1,\xi_2)$ mode, i.e., the excitation occurs within the diquark $(ss)$ or anti-diquark $(\bar{s}\bar{s})$.
These two states may have large decay rates into $\phi f_1(1285)$ and
$\phi f_1(1420)$ in an $S$ wave, or into $\eta \phi$ and $\eta' \phi$ in a $P$ wave through the quark rearrangements. These $0^{--}$ exotic states may be produced by the reactions $e^+e^-\to \eta^{(')}X\to \eta^{(')}\eta^{(')} \phi$ or $J/\psi\to \eta^{(')}\eta^{(')} \phi$.

\subsubsection{$2^{--}$ states}

There are three $2^{--}$ states, $T_{(ss\bar{s}\bar{s})2^{--}}(2446)$,
$T_{(ss\bar{s}\bar{s})2^{--}}(2657)$, and $T_{(ss\bar{s}\bar{s})2^{--}}(2907)$, predicted in the NRPQM.
Both the lowest mass state $T_{(ss\bar{s}\bar{s})2^{--}}(2446)$ and highest mass state
$T_{(ss\bar{s}\bar{s})2^{--}}(2907)$ are mixed states between the two color structures,
$|6\bar{6}\rangle_c$ and $|\bar{3}3\rangle_c$, and their orbital excitations are dominated by
the $(\xi_1,\xi_2)$ mode. The middle state $T_{(ss\bar{s}\bar{s})2^{--}}(2657)$
is dominated by the $^{3}P_{2^{--}(\bar{3}3)_{c}\left(\xi_{3}\right)}$ configuration, of which the orbital excitation occurs between the diquark $(ss)$ and anti-diquark $(\bar{s}\bar{s})$.
A corresponding state in the relativized
quark model~\cite{Lu:2019ira} has a mass of 2622 MeV, which is consistent with our prediction.
These $2^{--}$ states may easily decay into $\phi f_1(1285)$, $\phi f_1(1420)$, and $\phi f_2'(1525)$ in an $S$ wave, or into $\eta\phi$, $\eta'\phi$ in a $P$ wave through the quark rearrangements. They can also be searched for in $e^+e^-\to \eta^{(')}X\to \eta^{(')}\eta^{(')} \phi$ or vector charmonium decays such as $J/\psi\to \eta\eta\phi$.

\subsubsection{$3^{--}$ state}

There is only one $3^{--}$ state $T_{(ss\bar{s}\bar{s})3^{--}}(2719)$ predicted in the NRPQM.
This state has a pure color structure $|\bar{3}3\rangle_c$, and also
a pure orbital excitation between the diquark $(ss)$ and anti-diquark $(\bar{s}\bar{s})$.
Our predicted mass is about 60 MeV larger than that predicted by the relativized
quark model~\cite{Lu:2019ira} with a diquark approximation.
The $3^{--}$ states may easily decay into $\phi f_2'(1525)$ in an $S$ wave by the quark rearrangements. Since it has a high spin, it may be produced relatively easier in  $p\bar{p}$ or $pp$ collisions.

\section{Summary}\label{Summary}

In this work we calculate the mass spectra for the $1S$, $1P$ and $2S$-wave $T_{ss\bar{s}\bar{s}}$ states
in a nonrelativistic potential quark model without the often-adopted diquark-antidiquark approximation. The $1S$-wave ground states lie
in the mass range of $\sim 2.21-2.44$ GeV, while the $1P$- and $2S$-wave states scatter in a rather
wide mass range of $\sim 2.44-2.99$ GeV. For the $2S$-wave states, except for the
highest state $T_{(ss\bar{s}\bar{s})0^{++}}(3155)$ all the other states lie in a relatively narrow range of
$\sim 2.78-2.98$ GeV. We find that most of the physical states are mixed states with
different configurations.

For the $ss\bar{s}\bar{s}$ system it shows that both the kinetic energy $\langle T\rangle$ and the linear confinement potential $\langle V^{Lin}\rangle$
contribute a large positive value to the mass, while the Coulomb type potential $\langle V^{Coul}\rangle$
has a large cancellation with the these two terms. The spin-spin interaction $\langle V^{SS}\rangle$,
tensor potential $\langle V^{T}\rangle$, and/or the spin-orbit interaction term $\langle V^{LS}\rangle$
also have sizeable contributions to some configurations.

Some $T_{ss\bar{s}\bar{s}}$ states may have shown hints in experiment.
For instance, the observed decay modes and masses of $f_0(2200)$ and $f_2(2340)$ listed
in RPP~\cite{Tanabashi:2018oca} could be good candidates for the ground states
$T_{(ss\bar{s}\bar{s})0^{++}}(2218)$ and $T_{(ss\bar{s}\bar{s})2^{++}}(2378)$,
respectively. The newly observed $X(2500)$ at BESIII may be a candidate for the
lowest mass $1P$-wave $0^{-+}$ state $T_{(ss\bar{s}\bar{s})0^{-+}}(2481)$.
Another $0^{++}$ ground state
$T_{(ss\bar{s}\bar{s})0^{++}}(2440)$ may have shown signals in the $\phi\phi$ channel
at BESIII~\cite{Kozhevnikov:2019lmy,Ablikim:2016hlu}. Our calculation shows that $\phi(2170)$ may not favor
a vector state of $T_{ss\bar{s}\bar{s}}$, because of the much higher mass obtained in our model.

It should be stressed that as a flavor partner of $T_{cc\bar{c}\bar{c}}$, the $T_{ss\bar{s}\bar{s}}$ system may have very different dynamic features that need further studies. One crucial point is that the strange quark is rather light and the light flavor mixing effects could become non-negligible. It suggests that strong couplings between $T_{ss\bar{s}\bar{s}}$ and open strangeness channels could be sizeable. As a consequence, mixings between $T_{ss\bar{s}\bar{s}}$ and $T_{sq\bar{s}\bar{q}}$ would be inevitable. For an $S$-wave strong coupling, it may also lead to configuration mixings which can be interpreted as hadronic molecules for a near-threshold structure. In such a sense, this study can set up a reference on the basis of orthogonal states. More elaborated dynamics can be investigated by including the hadron interactions in the Hamiltonian. For states with exotic quantum numbers, experimental searches for their signals can be carried out at BESIII and Belle-II.

\section*{Acknowledgement}

The authors thank Wen-Biao Yan for useful discussions on the BESIII results. Helpful discussions with Atsushi Hosaka and Qi-Fang L\"{u} are greatly appreciated.
This work is supported by the National Natural Science Foundation of China (Grants Nos. 11775078, U1832173,
11705056, 11425525, and 11521505). Q.Z. is also supported in part, by the DFG and NSFC funds to the Sino-German CRC 110 ``Symmetries and the Emergence of Structure in QCD'' (NSFC Grant No. 12070131001), the Strategic Priority Research Program of Chinese Academy of Sciences (Grant No. XDB34030302), and National Key Basic Research Program of China under Contract No. 2015CB856700.

\bibliographystyle{unsrt}

\end{document}